\documentclass[iop]{emulateapj}

\usepackage{graphicx,amsmath,natbib,color,verbatim,subfigure}
\usepackage[normalem]{ulem}
\newcommand{\myemail}{rnilsson@amnh.org}
\newcommand\changed[1]{#1} % Set to normal

\slugcomment{{\bf Submitted to {\it Astrophysical Journal} 26 Jul 2015. Revised 27 Mar 2016, 26 Dec 2016, and 17 Feb 2017. Accepted 28 Feb 2017}}

\shorttitle{Project\,1640 Observations of Brown Dwarf GJ\,758\,B}
\shortauthors{Nilsson et al.}

\begin{document}

\title{Project\,1640 Observations of Brown Dwarf GJ\,758\,B: Near-Infrared Spectrum and Atmospheric Modeling}

% Main authors sorted by contribution
\author{R.~Nilsson\altaffilmark{1,2,3}}
\author{A.~Veicht\altaffilmark{1}}
\author{P.~A.~Giorla~Godfrey\altaffilmark{1,4,5}}
\author{E.~L.~Rice\altaffilmark{1,4,5}}
\author{J.~Aguilar\altaffilmark{6}}
\author{L.~Pueyo\altaffilmark{7}}
\author{L.~C.~Roberts,~Jr.\altaffilmark{8}}
\author{R.~Oppenheimer\altaffilmark{1}}
\author{D.~Brenner\altaffilmark{1}}
\author{S.~H.~Luszcz-Cook\altaffilmark{1}}

% Additional authors sorted alphabetically
\author{E.~Bacchus\altaffilmark{9}}
\author{C.~Beichman\altaffilmark{10,1}}
\author{R.~Burruss\altaffilmark{8}}
\author{E.~Cady\altaffilmark{8}}
\author{R.~Dekany\altaffilmark{11}}
\author{R.~Fergus\altaffilmark{12}}
\author{L.~Hillenbrand\altaffilmark{3}}
\author{S.~Hinkley\altaffilmark{13}}
\author{D.~King\altaffilmark{9}}
\author{T.~Lockhart\altaffilmark{8}}
\author{I.~R.~Parry\altaffilmark{9}}
\author{A.~Sivaramakrishnan\altaffilmark{7}}
\author{R.~Soummer\altaffilmark{7}}
\author{G.~Vasisht\altaffilmark{8}}
\author{C.~Zhai\altaffilmark{8}}
\author{N.~T.~Zimmerman\altaffilmark{7}}

\altaffiltext{1}{Astrophysics Department, American Museum of Natural History, Central Park West at 79th Street, New York, NY 10024, USA; {\myemail}.}
\altaffiltext{2}{Department of Astronomy, Stockholm University, AlbaNova University Center, Roslagstullsbacken 21, SE-106 91 Stockholm, Sweden}
\altaffiltext{3}{Department of Astronomy, California Institute of Technology, 1200 E. California Blvd, MC 249-17, Pasadena, CA 91125 USA}
\altaffiltext{4}{Physics Program, The Graduate Center, City University of New York, New York, NY 10016, USA}
\altaffiltext{5}{Department of Engineering Science \& Physics, College of Staten Island, 2800 Victory Blvd., Staten Island, NY 10314 USA}
\altaffiltext{6}{Department of Physics and Astronomy, The Johns Hopkins University, Baltimore, MD 21218 USA}
\altaffiltext{7}{Space Telescope Science Institute, 3700 San Martin Drive, Baltimore, MD 21218 USA}
\altaffiltext{8}{Jet Propulsion Laboratory, California Institute of Technology, 4800 Oak Grove Dr., Pasadena CA 91109 USA}
\altaffiltext{9}{Institute of Astronomy, Cambridge University, Madingley Road, Cambridge CB3 0HA United Kingdom}
\altaffiltext{10}{NASA Exoplanet Science Institute, California Institute of Technology, Pasadena, CA 91125 USA}
\altaffiltext{11}{Caltech Optical Observatories, California Institute of Technology, Pasadena, CA 91125 USA}
\altaffiltext{12}{Department of Computer Science, Courant Institute of Mathematical Sciences, New York University, 715 Broadway, New York, NY 10003 USA}
\altaffiltext{13}{School of Physics, University of Exeter, Stocker Road, Exeter, EX4 4QL}
%\altaffiltext{13}{Department of Mechanical and Aerospace Engineering, Princeton University, Princeton, NJ 08544, USA}

\begin{abstract}
The nearby Sun-like star GJ\,758 hosts a cold substellar companion, GJ\,758\,B, at a projected separation of $\lesssim 30$\,au, previously detected in high-contrast multi-band photometric observations. In order to better constrain the companion's physical characteristics, we acquired the first low-resolution ($R \sim 50$) near-infrared spectrum of it using the high-contrast hyperspectral imaging instrument Project\,1640 on Palomar Observatory's 5\,m Hale telescope. %Aims \\
We obtained simultaneous images in 32 wavelength channels covering the $Y$, $J$, and $H$ bands ($\sim$952--1770\,nm), and used data processing techniques based on principal component analysis to efficiently subtract chromatic background speckle-noise.  % Methods \\
GJ\,758\,B was detected in four epochs during 2013 and 2014. Basic astrometric measurements confirm its apparent northwest trajectory relative to the primary star, with no clear signs of orbital curvature. Spectra of SpeX/IRTF observed T~dwarfs were compared to the combined spectrum of GJ\,758\,B, with ${\chi}^2$ minimization suggesting a best fit for spectral type {T7.0$\pm$1.0}, but with a shallow minimum over T5--T8. Fitting of synthetic spectra from the BT-Settl13 model atmospheres gives an effective temperature {$T_{\text{eff}} = 741 \pm 25$\,K} and surface gravity {$\log g = 4.3 \pm 0.5$\,dex} (cgs). Our derived best-fit spectral type and effective temperature from modeling of the low-resolution spectrum suggest a slightly earlier and hotter companion than previous findings from photometric data, but do not rule out current results, and confirm GJ\,758\,B as one of the coolest sub-stellar companions to a Sun-like star to date. % Results \\
\end{abstract}

\keywords{instrumentation: adaptive optics -- instrumentation: spectrographs -- planets and satellites: detection -- brown dwarfs -- stars: individual (\object{GJ\,758}) -- techniques: high angular resolution}

\section{Introduction} \label{sec:intro}

Techniques for high-contrast imaging have now matured to a level where direct detections of substellar companions to nearby stars are becoming frequent. Adaptive optics combined with coronagraphy, and sophisticated software speckle-reduction techniques using angular differential imaging and spectral differential imaging has revealed low mass-ratio companions at $10$--$100$\,au around A\,stars \citep{Lagrange2010,Marois2010,Rameau2013}, and high mass-ratio companions at wider separations around later type stars \citep{Lafreniere2008,Currie2014a}.\footnote{This should not be taken to imply that high mass-ratio companions to early type stars do not exist, as recently shown in aperture masking interferometry observations by \citet{Hinkley2015a}.} Distinguishing between different formation scenarios, primarily core accretion \citep[e.g.,][]{Pollack1996} or gravitational instability \citep[e.g.,][]{Boss2011}, requires tight constraints on the orbital and physical parameters of these systems. It should be noted that direct imaging is still restricted to relatively large separations (beyond tens of au) and high mass ($>$\,$M_\mathrm{Jup}$) and/or young companions. 

Out of the current two dozen or so directly imaged substellar companions, some have still only been observed in thermal emission in a few broad near-infrared (near-IR) photometric bands. However, instruments like \emph{Project\,1640} \citep[P1640;][]{Oppenheimer2012} at the Palomar Hale telescope, \emph{Gemini Planet Imager} \citep{Macintosh2014} at Gemini South, and \emph{SPHERE} \citep{Beuzit2006} at the Very Large Telescope, are now able to simultaneously image and obtain low-resolution spectra using chromatic speckle suppression, allowing improved atmospheric characterization of gas-giant exoplanets and companion BDs.

\object{GJ\,758} (\object{HIP\,95319}, \object{HD\,182488}) is a Sun-like (spectral type G8\,V) star located 15.76\,pc away \citep{VanLeeuwen2007}, that has a substellar companion (GJ\,758\,B) at a projected separation of $\sim$\,30\,au, detected in $H$ band Subaru/HiCIAO imaging by \cite{Thalmann2009}, confirmed in $L^{\prime}$ band MMT/Clio imaging by \cite{Currie2010}, and followed up by multi-band ($J$, $H$, $K_{c}$, $L^{\prime}$, $M$, and narrow band $CH4S$ and $CH4L$ filters) Subaru/HiCIAO, Gemini/NIRI, and Keck/NIRC2 imaging by \cite{Janson2011b}.\footnote{During the referee process of this paper, additional results from observations with VLT/SPHERE were published by \citet{Vigan2016}.} Models presented in those papers have suggested an effective surface temperature of $\sim$600\,K, making it a T8--T9 dwarf, the coldest imaged companion of a Sun-like star, and one of the most important ``planet-like" objects accessible for detailed study due to its proximity. The derived companion mass depends chiefly on its age, which, using any current method of age determination for main-sequence stars, remains highly uncertain. A wide range of possible ages, from 0.7--8.7\,Gyr, have been derived, suggesting a companion mass ranging from $10\,M_\text{Jup}$ (exoplanet region) to $40\,M_\mathrm{Jup}$ (BD region). While \citet{Takeda2007} derived an age of 0.7\,Gyr based on isochronal fits, \citet{Valenti2005a} and \citet{Holmberg2009} suggested several Gyr using the same method. Together with a non-detection of lithium in the atmosphere of GJ\,758, \citet{Thalmann2009} decided to exclude the lower age in favor of activity and rotation based ages of $\sim$\,5--9\,Gyr \citep{Barnes2007,Mamajek2008,Thalmann2009}, narrowing the estimated mass range to 30--$40\,M_\mathrm{Jup}$.

Although spectra have been obtained of many field BDs, including late T~dwarfs, their even less constrained ages and metallicities make them hard to confidently model. Tightly constraining the physical properties of T~dwarf \emph{companions} around sun-like stars will greatly contribute to our understanding of BDs in general, and place them in relation to planets and planetary system formation scenarios. 

In this paper we present high-contrast imaging observations of \object{GJ\,758}, using Hale/P1640, where we obtain the first low-resolution near-IR spectrum of the B companion, at four epochs, and model the companion's temperature, surface gravity, and spectral type. We also confirm the presence of methane in its atmosphere, as previously suggested from multi-band photometry \citep{Janson2011b}. A follow-up paper (J.~Aguilar et al. 2017, in preparation) will present more detailed astrometric analysis and orbital simulations, using our new detection epochs and previous observations.

\section{Observations} \label{sec:observations}

Palomar Observatory's 5.1\,m Hale telescope is equipped with PALM-3000 \citep[P3k;][]{Dekany2013a}, a woofer-tweeter AO system with a 3388 actuator high-order deformable mirror (DM) and a 241 actuator low-order DM, which was used in conjunction with the P1640 instrument \citep{Oppenheimer2012}. P3k can, with recent hardware and software upgrades, reach Strehl ratios of 92\% in the $K$ band (100\,nm rms wavefront error, corresponding to 0.86\%, 0.76\%, 0.68\% Strehl ratios in $H$, $J$ and $Y$, respectively), on bright stars and in good seeing conditions \citep[][and R.~Burruss private communication 2016]{Burruss2014}. P1640 combines an apodized pupil Lyot coronagraph \citep{Hinkley2009,Soummer2009} with an internal calibration system \citep[CAL;][]{Zhai2012,Cady2013,Vasisht2014} for wavefront sensing and correction of residual phase and amplitude distortions at the occulter, and an integral-field spectrograph (IFS) with a $200\times200$ lenslet array sampling the focal plane. Details on the instrument can be found in \cite{Oppenheimer2012} and \cite{Hinkley2011}. The covered wavelength range is 969--1797\,nm, encompassing the near-IR $Y$, $J$, and $H$ bands in 32 channels, at a spectral resolution of $\Delta\lambda = 26.7$\,nm. The total field-of-view (FOV) is approximately $3{\farcs}8 \times 3{\farcs}8$.

\object{GJ\,758} was observed at an airmass of 1.00--1.05 at five different occasions\footnote{UT dates are given as year-month, year-month-day (in compliance with ISO\,8601) throughout the paper, and as Julian days and Besselian years in Table~\ref{tab:astrometry}.}, from 2012-06 to 2014-09, as presented in Table~\ref{tab:obs}. Observing conditions varied, with the 2013-10 data taken at $\sim$\,$1\farcs1$ seeing (in $V$ band), while 2012-06 data were obtained at a seeing far above the P1640 limiting $\sim$\,$1\farcs35$ value for specified performance, and the 2014-06 data suffering from significant ``mirror seeing'' from large temperature gradients above the telescope main mirror. After pointing the telescope to the star, and making an initial AO tune-up, we used an internal white-light source to iterate on low- and high-order wavefront corrections with CAL. Final rms wavefront errors were on the order of 10\,nm. Reference astrometric spots, for locating the stellar center behind the occulting mask \citep{Marois2006,Sivaramakrishnan2006}, were introduced by applying a sinusoidal pattern on the DM. At each observing occasion, we obtained 3--10 long (3--6\,minutes), occulted exposures, as well as 3--10 short (1.5--3.0\,s), unocculted, ``core'' exposures, of the star. We also observed spectral standard stars for spectral calibration, and binary standards for plate-scale and position angle (PA) calibration (see Section~\ref{sec:datared}).  

\begin{deluxetable*}{lccccc}
\tabletypesize{\scriptsize}
\tablecaption{Observations of GJ\,758 with Project\,1640}
\tablewidth{0pt}
\tablehead{\colhead{Date} & \colhead{Julian Date} & \colhead{$N_{\text{exp}} \times t_{\text{exp}}$} & \colhead{Estimated Seeing\tablenotemark{a}} & \colhead{rms-WFE\tablenotemark{b}} \\ \colhead{(UT)} & \colhead{(days)} & \colhead{(s)} & \colhead{($\arcsec$)} & \colhead{(nm)}}
\startdata
2012-06-17 & 2456095.8840277782 & $3 \times 366.6$ & 1.5 & ... \\
2013-07-21 & 2456494.8199421302 & $10 \times 185.9$ & ... & 7.5 \\
2013-10-18 & 2456583.5734722228 & $5 \times 185.9$, $6 \times 278.8$ & 1.1 & 4.5 \\
2014-06-10 & 2456818.9438657411 & $7 \times 371.8$ & \phantom{$^d$}1.0\tablenotemark{d} & 9.0 \\
2014-09-07 & 2456907.6952893524 & $8 \times 371.8$ & 1.4 & 9.0
\enddata
\tablenotetext{a}{Mean FWHM of PSF in $V$ band as recorded by Palomar 18\,inch seeing monitor observing Polaris. Not recorded on 2013-07-21.}
\tablenotetext{b}{High-order wavefront error from final CAL iteration. Not recorded on 2012-06-17.}
\tablenotetext{c}{Effective seeing was closer to $2\farcs0$ due to high temperature gradient, with the telescope mirror being a few degrees warmer than the outside air.}
\label{tab:obs}
\end{deluxetable*}

\section{Data Reduction} \label{sec:datared}

Descriptions of the overall P1640 data reduction procedure, as well as details of most individual pipeline modules, have been presented elsewhere \citep[see, e.g.,][]{Oppenheimer2013}, but for completeness we summarize the steps below.

\subsection{Cube extraction} \label{sec:datared:pcxp}
The raw images, consisting of nearly 40,000 tightly packed spectra, each covering roughly $32\times3$ pixels on the detector, are extracted and converted into data cubes (measuring $250\times250\times32$ in $x \times y \times \lambda$, where $x$ and $y$ are the number of pixels in the first and second dimension, and $\lambda$ is the number of wavelength channels in the third dimension) by the \emph{P1640 Data Cube Extraction Pipeline} \citep[PCXP;][]{Zimmerman2011}. PCXP uses the location of spectra (dots) from laser exposures at 1310 and 1550\,nm, together with sky flats, to create a focal plane solution that maps individual spectra to lenslets and corresponding boxel positions. Wavelength channels in which telluric atmospheric water absorption lines are strong (mainly channels 8, and 17--18) display some degree of cross-talk from adjacent spectra, and so does the first and last channel where our sensitivity falls steeply. Estimated errors at those wavelengths are consequently much larger than in the rest of the spectrum. Although those channels could have been omitted in the final analysis, we include them with corresponding errors in Section~\ref{sec:analysis} since trimming the spectra did not significantly change the results.

\subsection{Dispersion correction} \label{sec:datared:cacs}
Using the four astrometric reference spots, we determine the star's centroid position in each wavelength channel, and track both atmospheric and instrumental dispersion shifts through the cube. The image slices are then shifted in $x$ and $y$ with sub-pixel accuracy to align and center the star at $(x,y) = (126,126)$ in each cube (the cubes are later padded with a 251st row and column of zeros to place the center of the star in the center pixel). As the radial position of reference spots and speckle noise is wavelength dependent, we can also derive a radial scaling relation, scale each image, and use cross-correlation for even finer sub-pixel image registration. This is all handled in the \emph{Cube Alignment Centering and Stacking} (CACS, R.~Nilsson et al.\ 2017, in preparation) pipeline module, which gives a final image registration accuracy of $\sim$\,0.2 pixel (1-$\sigma$ deviation), and saves a hypercube ($x \times y \times \lambda \times N_{\text{exp}}$, where $N_\text{exp}$ is the number of exposures) for our speckle suppression algorithms. Aperture photometry of the reference spots also allow us to determine the $\lambda$ location of two telluric water absorption bands in our spectra (at 1100 and 1380\,nm), {which together with knowledge of the instrument filter bandpass edges can be used to derive the shift and stretch function that aligns} the cubes in the wavelength dimension.

\subsection{Speckle suppression} \label{sec:datared:s4d}
The wavelength-dependent speckle noise was modeled and subtracted using both KLIP \citep{Soummer2012} and S4 \citep{Fergus2014}, to check for consistency. Both algorithms employ principal component analysis (PCA), but with component decomposition performed in different dimensions. Karheunen--Loeve Image Projection is used to decompose the speckle noise pattern into its principal components (PCs) by projecting companion-free regions of an image onto a Karheunen--Loeve basis. The algorithms then use the most prominent set of PCs to forward model speckle noise in the image to remove flux associated with the speckles but not with a putative companion.

\subsection{Point-source identification} \label{sec:datared:planetfinder}
Speckle reduced (residual) images in each channel and cube were filtered by convolving them with a model point-spread function (PSF) core to reduce noise, producing a ``detection map'' in which we could search for companion signals \citep{Fergus2014}. We also produced signal-to-noise (S/N) maps by local calculation of the noise in each area of the residual image, using two different methods: (1) noise calculation in concentric annuli around the star (as normally done in high-contrast coronagraphic images with radially decreasing noise profiles), and (2) calculating the standard deviation of count levels in $20 \times 20$ pixel boxes around each pixel, with the inner $11 \times 11$ pixels left out. With the moving-box method we get an estimate of the increased noise along the astrometric spot trails, in regions that are cut out in the first method in order not to suppress overall S/N. The difference in calculated S/N between the two methods turns out to be less than {10\% in areas outside the spot trails}. Significant peaks were found by $\sigma$-clipping and searching for regional maxima. Detected peaks were weighted by S/N and summed up for each location, producing a list of the locations with the strongest peaks. At these locations we extracted raw mean spectra to compare with a range of reasonable companion spectra, and to exclude locations with a raw spectrum dominated only by bright peaks in the noisy water absorption channels. We also excluded peaks along the radial paths of the four astrometric reference spots.

\subsection{Spectral extraction} \label{sec:datared:s4s}
After locating the centroid pixel of a detected point-source in the residual cubes, we run S4's spectral extraction code, which again performs speckle-suppression on the aligned and centered hypercube, but this time with local optimization in the area where the suspected companion is located. By varying the model parameters (such as the size of test zone, $\Delta\theta$ and ${\Delta}r$, and the number of PCs, $N_{\text{PC}}$, used), and measuring the change of the extracted candidate companion spectrum, as well as of a number of (typically 50) background points and fake source insertions at the same radius, the parameters that minimize the noise and most faithfully retrieve the fake source spectrum can be determined. This ensures that the speckle noise is optimally modeled and subtracted, without attenuating the companion flux by overfitting the data. {A brief justification of chosen optimal S4 spectral extraction parameters is presented in Section~\ref{sec:pc_opt}. The results agree with the more rigorous treatment demonstrated in A.~Veicht et al.\ (2017, in preparation).}

Note that in the spectral extraction we also simultaneously fit both the speckle and companion models to the observed data, and the spectrum of the companion model is being jointly estimated along with the PCA coefficients of the companion model. The S/N of an extracted companion spectrum will thus in general be higher than for the original detection signal. {The total gain in S/N from the full optimization procedure is about a factor of two in $J$ and $H$ compared to the initial detection image.} Also note that the errors in the extracted companion spectrum are derived from the extracted fake insertions at the same projected radius, corresponding to the standard method of S/N determination in high-contrast imaging. More information about the S4 and KLIP spectral extractions procedure can be found in the appendix of \cite{Oppenheimer2013}, and a detailed analysis of S4's spectral extraction stability and error estimation is given in A.~Veicht et al.\ (2017, in preparation).

\subsubsection{Spectral calibration} \label{sec:datared:calib}
Each night, we obtained unocculted exposures of several calibrator stars of well-known spectral types with spectra available in the IRTF catalog of IR spectral standards \citep{Cushing2008,Rayner2009}. They were observed just before or after the occulted GJ\,758 exposures, at similar airmass. Although core exposures of the G8\,V primary could in principle have been used, they all saturated the detector in our shortest exposure setting. Calibrator PSF cores were fitted and shifted on a sub-pixel level to correct for atmospheric and instrument induced dispersion, and center the stars through the image cube. Their integrated count levels per wavelength channel, $S_\text{cal}(\lambda_\text{P1640})$, were found using aperture photometry, with a circular aperture enclosing the outermost visible Airy ring, and surrounding annulus for background calculation and subtraction. The IRTF template spectrum, $S_\text{temp}(\lambda_\text{IRTF})$, corresponding to the spectral type of the observed calibrator, was degraded to P1640 spectral resolution by convolving with a Gaussian having a full width at half maximum (FWHM) equal to the resolution of our instrument, and then rescaling to match the integrated flux in our wavelength range. The Spectral Response Function (SRF), which the extracted companion flux is divided with to correct for atmospheric transmission and instrument sensitivity, is given by the ratio of the extracted calibrator flux to the down-sampled template spectrum: $R(\lambda_\text{P1640}) = S_\text{cal}(\lambda_\text{P1640}) / S_\text{temp}(\lambda_\text{P1640})$.

To validate the SRF used to correct each hypercube and determine the SRF airmass dependence, the chosen SRFs were compared to a sequence of calibrator exposures obtained over several observing runs. As an additional consistency check on the shape and wavelength solution of each SRF, we compared them with the SRFs derived from the astrometric grid spots in the \object{GJ\,758} occulted data. Photometry of the latter has to be multiplied by $\lambda^2$ due to the chromaticity of the spot-inducing DM ripple, but can otherwise be used for simultaneous spectral calibration \citep{Sivaramakrishnan2006}, assuming that they are bright enough for a high S/N, and that the star's infrared spectrum and/or spectral type is well known.

Our wavelength accuracy depends on the cube extraction with PCXP \citep{Zimmerman2011}, which uses the focal-plane solution derived from sky flats and monochromatic light exposures at 1310\,nm and 1150\,nm, and which can often be up to one channel off in either direction. However, as previously mentioned (see Section~\ref{sec:datared:cacs}), anchoring and scaling the wavelength axis using the two telluric water absorption bands (at 1100.1 and 1380.1\,nm){, and the filter edges,} brings the accuracy down to below $\pm$\,26.4\,nm (the width of a channel).

\section{Results} \label{sec:results}

In this section we present resulting, PCA optimized data products from the speckle-suppression code: residual hypercubes ($x \times y \times \lambda \times N_{\text{exp}} \times N_{\text{PC}}$) and extracted companion spectra. GJ\,758\,B is weakly detected in four out of five epochs (including one marginal detection, see Table~\ref{tab:obs}). KLIP revealed the companion in data from 2013-10 and 2014-09, and was used to verify the results of S4 in those two epochs. Our main results are thus based on S4 produced data products, which are our primary focus below.

\subsection{Residual images and S/N maps} \label{sec:results:maps}

\begin{figure*}[htb]
\center
\resizebox{1.0\hsize}{!}{\includegraphics[width=.5\linewidth]{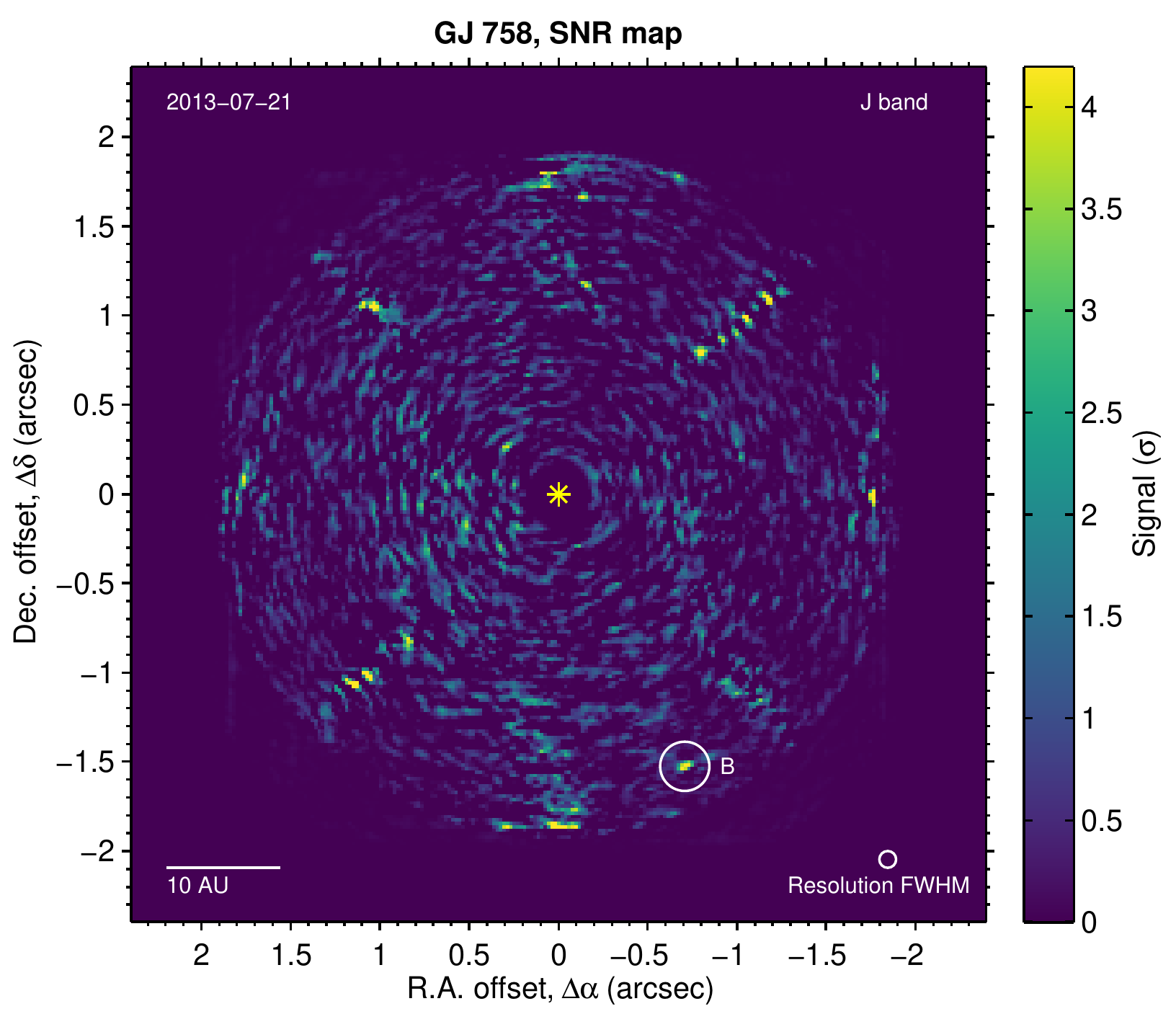} \includegraphics[width=.5\linewidth]{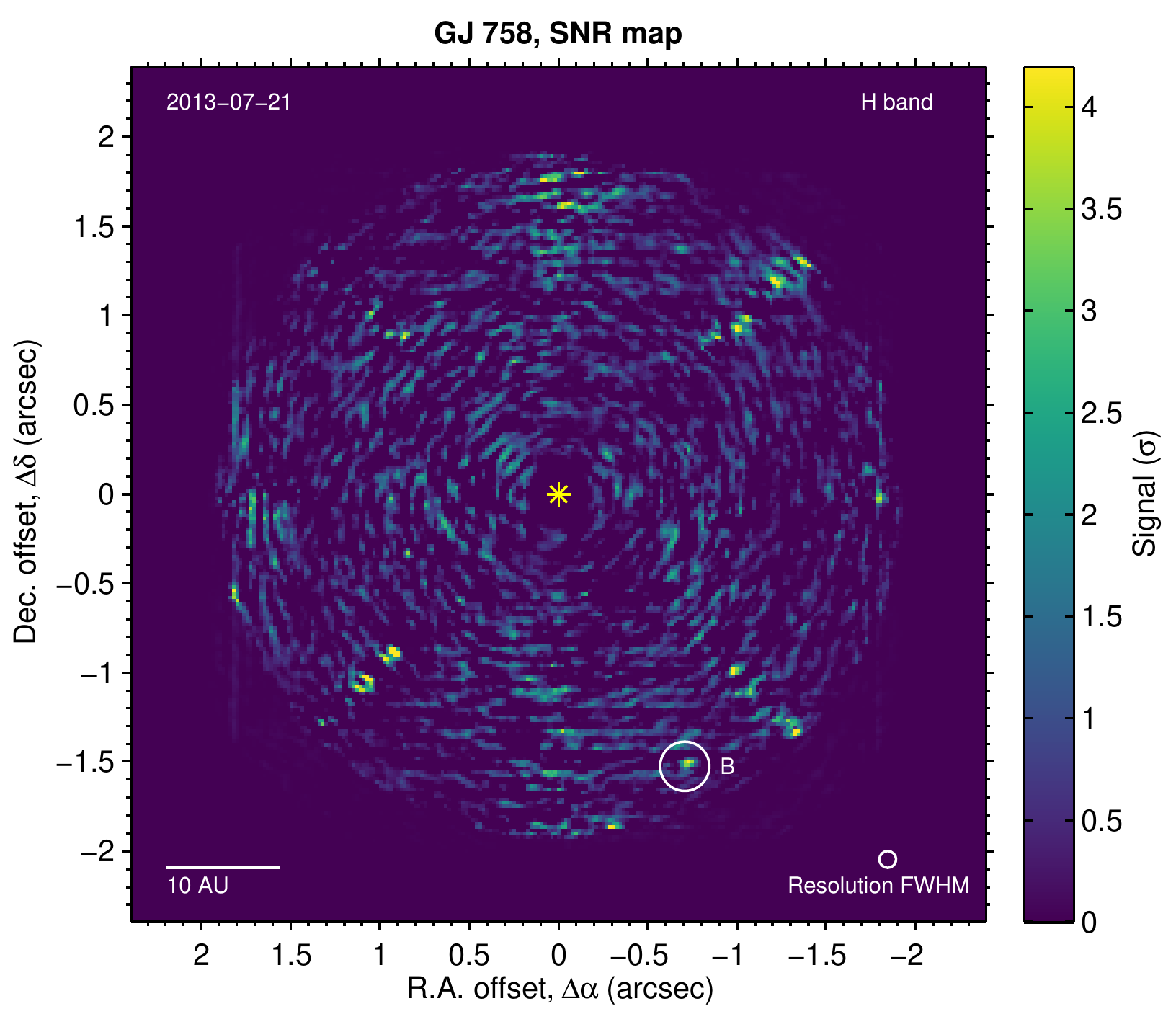}}
\caption{{S/N maps of GJ\,758 showing $J$ and $H$ images separately for epoch 2013-07. The images are from first-pass S4 processing with $N_{\textrm{PC}}$ = 50--110., displaying somewhat weaker detection significance than the locally optimized spectral extraction presented in Table~\ref{tab:detection_levels} and following figures. North and east orientation are up and left, respectively. The scale of 10\,au at the distance of the star is represented by a white line to the lower left in each image, while the circle to the lower right shows the spatial resolution, FWHM = 0$\farcs$0958, measured from the PSFs of multiple core exposures. Both maps use the same linear color scale.}}
\label{fig:GJ758_SNRmaps1}
\end{figure*}

\begin{figure*}[htb]
\center
\resizebox{1.0\hsize}{!}{\includegraphics[width=.5\linewidth]{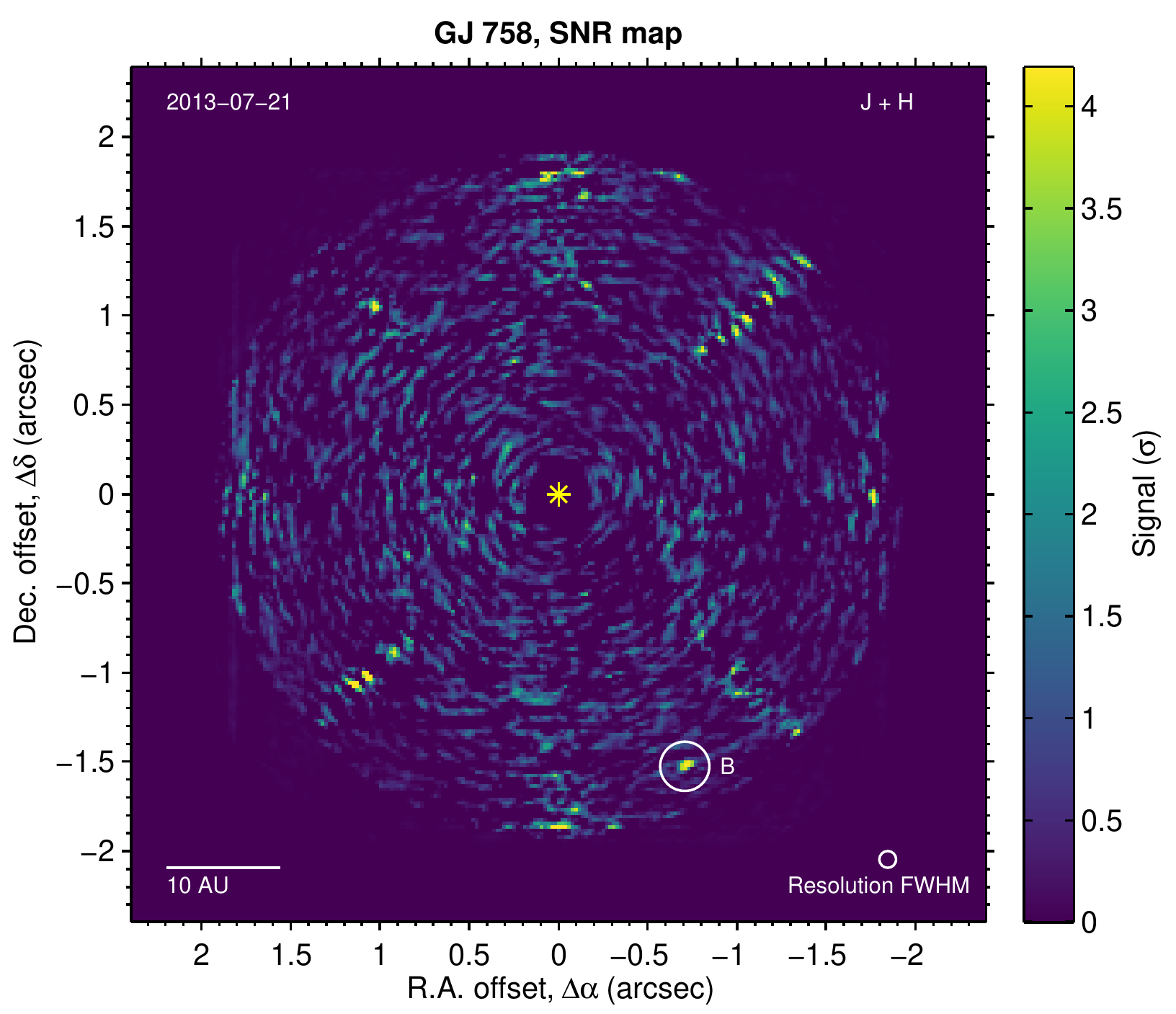} \includegraphics[width=.5\linewidth]{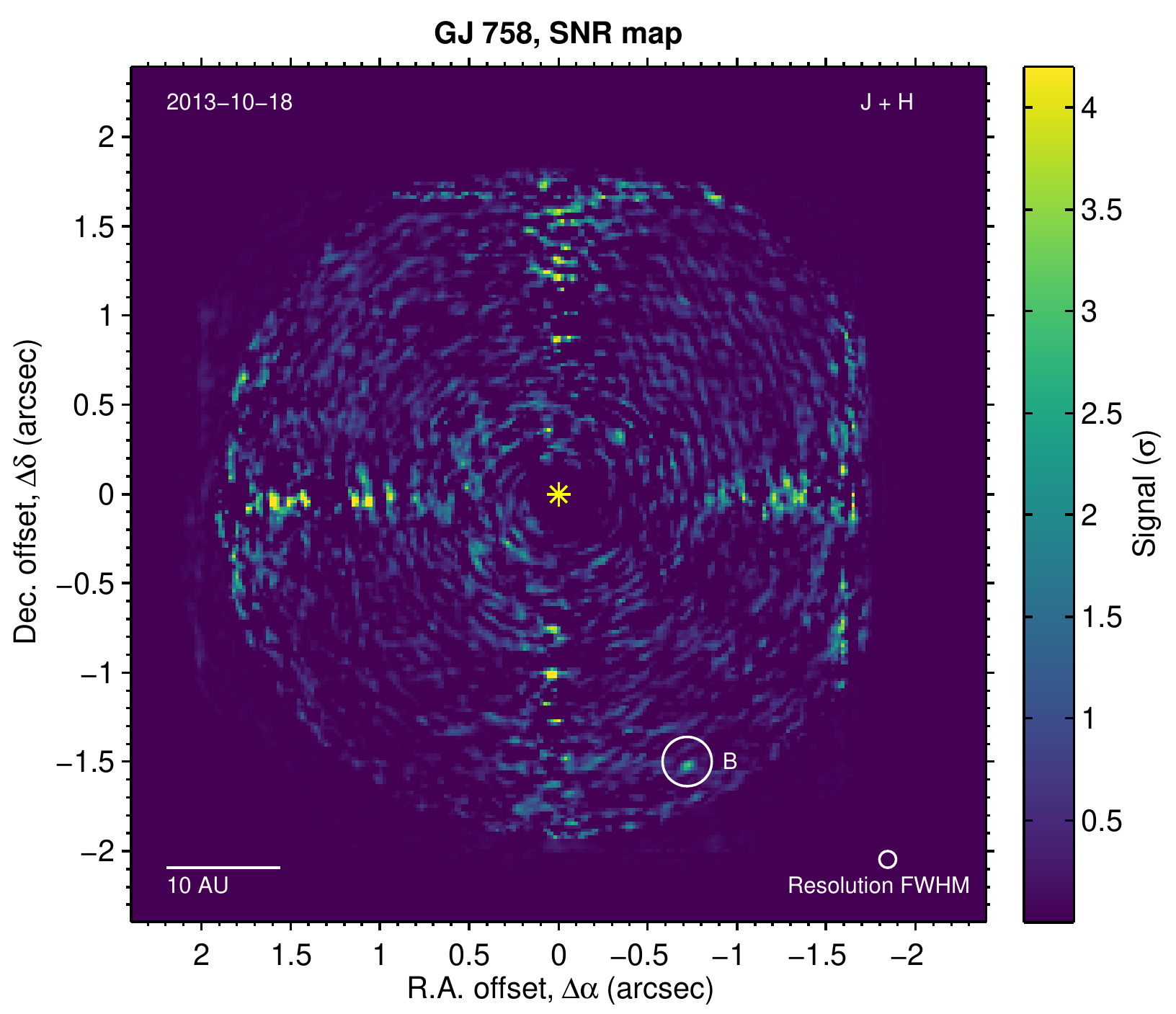}} \\
\resizebox{1.0\hsize}{!}{\includegraphics[width=.5\linewidth]{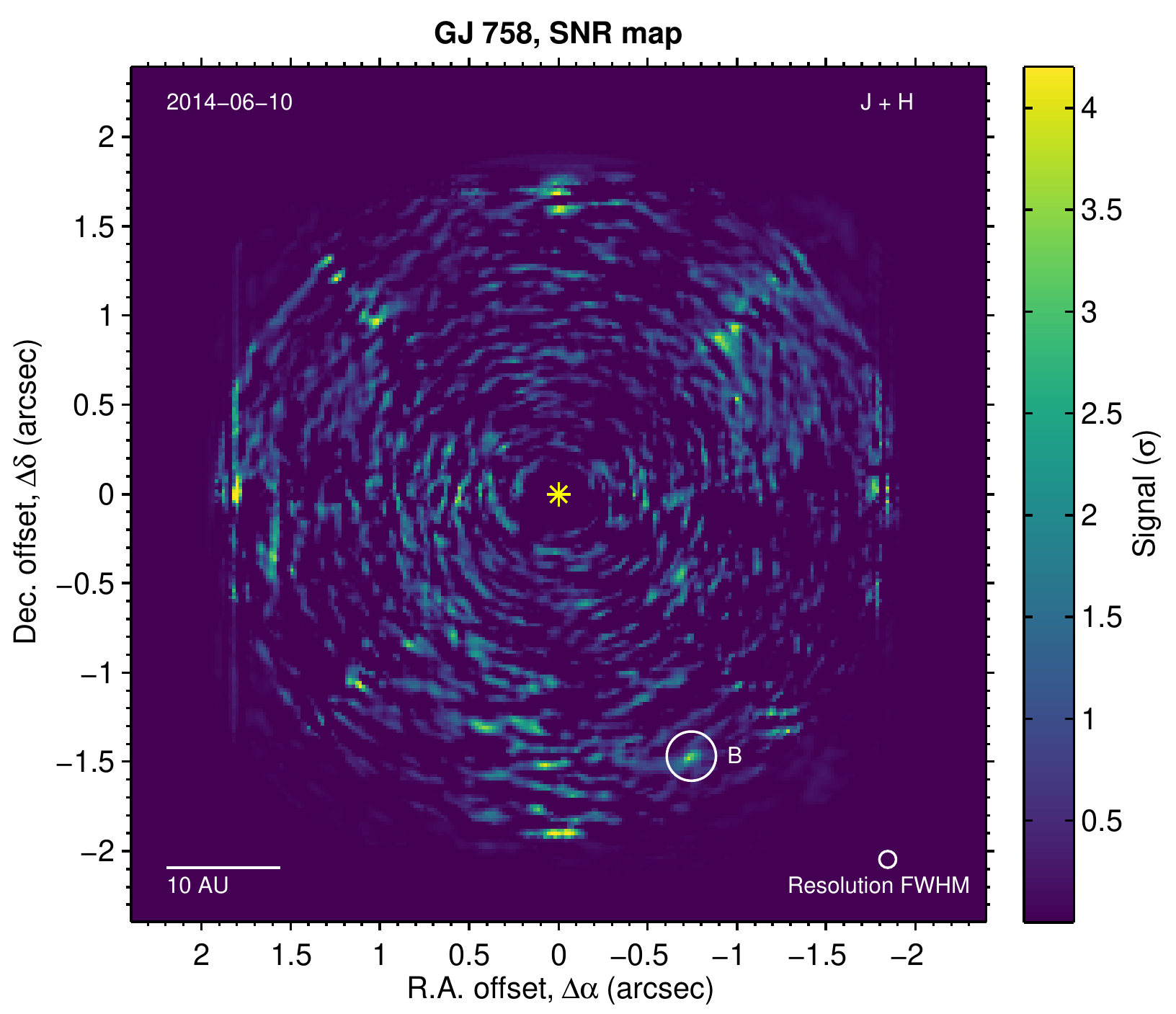} \includegraphics[width=.5\linewidth]{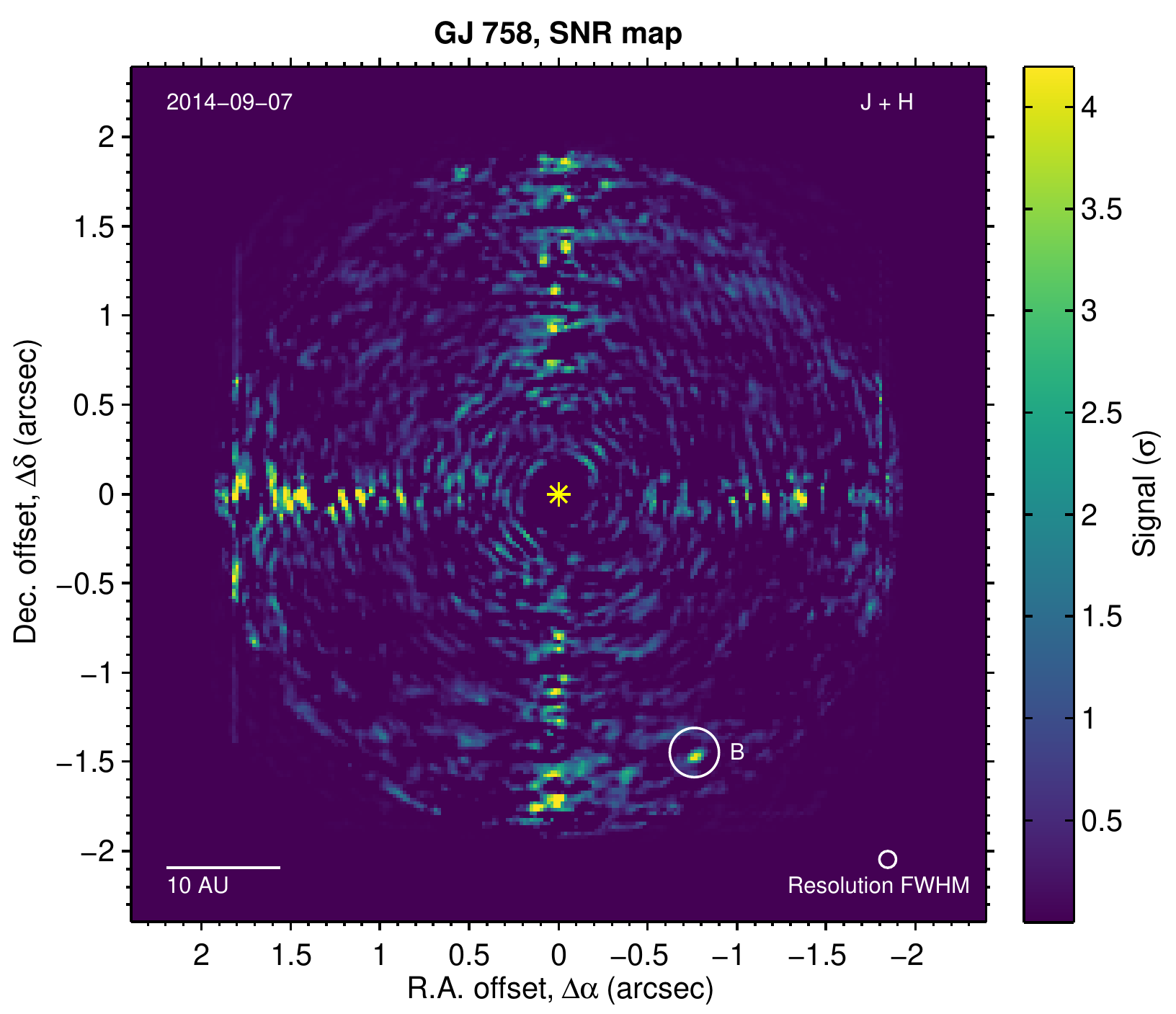}}
\caption{{S/N maps of GJ\,758 showing combined images from wavelength channels covering the $J$ and $H$ band for all four epochs. The images are from first-pass S4 processing with $N_{\textrm{PC}}$ = 50--110.}}
\label{fig:GJ758_SNRmaps2}
\end{figure*}

Speckle-reduced image cubes were examined, both using the source detection method outlined in Section~\ref{sec:datared:planetfinder} and by eye, to locate potential companion peaks for spectral extraction. At the expected location of GJ\,758\,B \citep[extrapolated from astrometry given in][]{Janson2011b} we find significant ($>3\sigma$) flux peaks in three epochs, and a marginal ($>2.5\sigma$) detection in one epoch (see Table~\ref{tab:obs} for detection levels), in multiple wavelength channels in the $J$ and $H$ bands. Although we are measuring some flux in $Y$, it is at low statistical significance and can {only be considered a marginal detection}. \changed{In Fig.~\ref{fig:GJ758_SNRmaps1} we show S/N maps for the $J$ and $H$ bands, separately, for the 2013-07-21 epoch, averaged over all exposures and over the number of PCs that give the highest S/N. Combined $J$ and $H$ S/N maps for all four epochs are shown in Fig.~\ref{fig:GJ758_SNRmaps2}}. Due to computational limitations, we ran S4 with a fixed $\Delta\theta = 3$ pixels in the detection phase, which gave highest companion S/N \changed{in PC range 50--110}. Note that this is not the same optimal number of PCs as determined for the locally optimized spectral extraction (see Section~\ref{sec:pc_opt} and Section~\ref{sec:res:spectrum}), which usually gives an optimum at $\Delta\theta = 5$, and $N_{\text{PC}}= 100$--250.

Although several other peaks in individual S/N maps seem to reach $>$3-$\sigma$ levels, all of them were discarded based on them either moving radially through consecutive wavelength channels (indicative of being a residual speckle), or being located along the trails of the four astrometric spots. Bright spot residuals are visible diagonally in the images from 2013-07 and 2014-06, and horizontally/vertically in images from 2013-10 and 2014-09 (Fig.~\ref{fig:GJ758_SNRmaps2}). We do not detect the faint background star previously found by \citet{Thalmann2009} and \citet{Janson2011b} within our FOV. Due to the high proper motion of \object{GJ\,758}, the star should have moved considerably to the southwest, but still remain within our FOV (see Section~\ref{sec:res:astrometry} and Fig.~\ref{fig:astrometry}). It is likely too faint to be seen in our images.

As the contrast in our observations could not be reliably calculated due to saturation of the primary in core exposures, we instead estimate reached contrast levels by comparing signal strengths to the absolute magnitudes of the companion \citep[from][]{Janson2011b} and the primary. GJ\,758\,B has $M_{J} = 17.58$ and $M_{H} = 18.16$, and the parent star has $M_{J} = 4.38$ and $M_{H} = 3.76$. This indicates 3-$\sigma$ contrasts of $2 \times 10^{-6}$ and $7 \times 10^{-7}$ in $J$ and $H$ respectively, reached at an angular separation of $1{\farcs}6$--$1{\farcs}7$.

\subsection{Astrometry} \label{sec:res:astrometry}

Determining the location of a companion relative its host star with high precision in coronagraphic data is notoriously difficult. First, finding the position of the stellar core in the focal plane, hidden behind the coronagraphic mask, poses challenges. Second, we have to measure the precise location of a faint point-source signal in an image littered with residual speckles. For IFUs in particular, there are additional concerns when calibrating the field-distortion, plate-scale, and detector orientation.

On each observing occasion, we observed a sample of binaries (HIP\,72447 for 2013-07; HD\,13594, HIP\,34860 and HIP\,97222 for 2013-10; HIP\,88745 and HIP\,107354 for 2014-06; and HIP\,10403, HIP\,12619, HD\,3304, HD\,11803, HIP\,25826, HIP\,97222, and HIP\,88745 for 2014-09) chosen from the Sixth Catalog of Orbits of Visual Binary Stars \citep{Hartkopf2011} to have well-determined orbital parameters. Several exposures of each binary were obtained, at different positions on the detector to examine effects of field-distortion. The plate scale and absolute north orientation of the detector were found to be very stable, and we calculated values for the different epochs all agreeing to within 1-$\sigma$ errors. We used the derived mean plate scale, $19.16 \pm 0.18$\,mas/pixel, and PA offset of $-72{\fdg}43 \pm 1{\fdg}06$ for all epochs. It should be noted that several of the binary calibrators lacked cataloged error estimates on their orbits, which means that the uncertainties on the plate scale and PA offset may be higher than stated.

\begin{figure}[htb]
\includegraphics[width=0.5\textwidth]{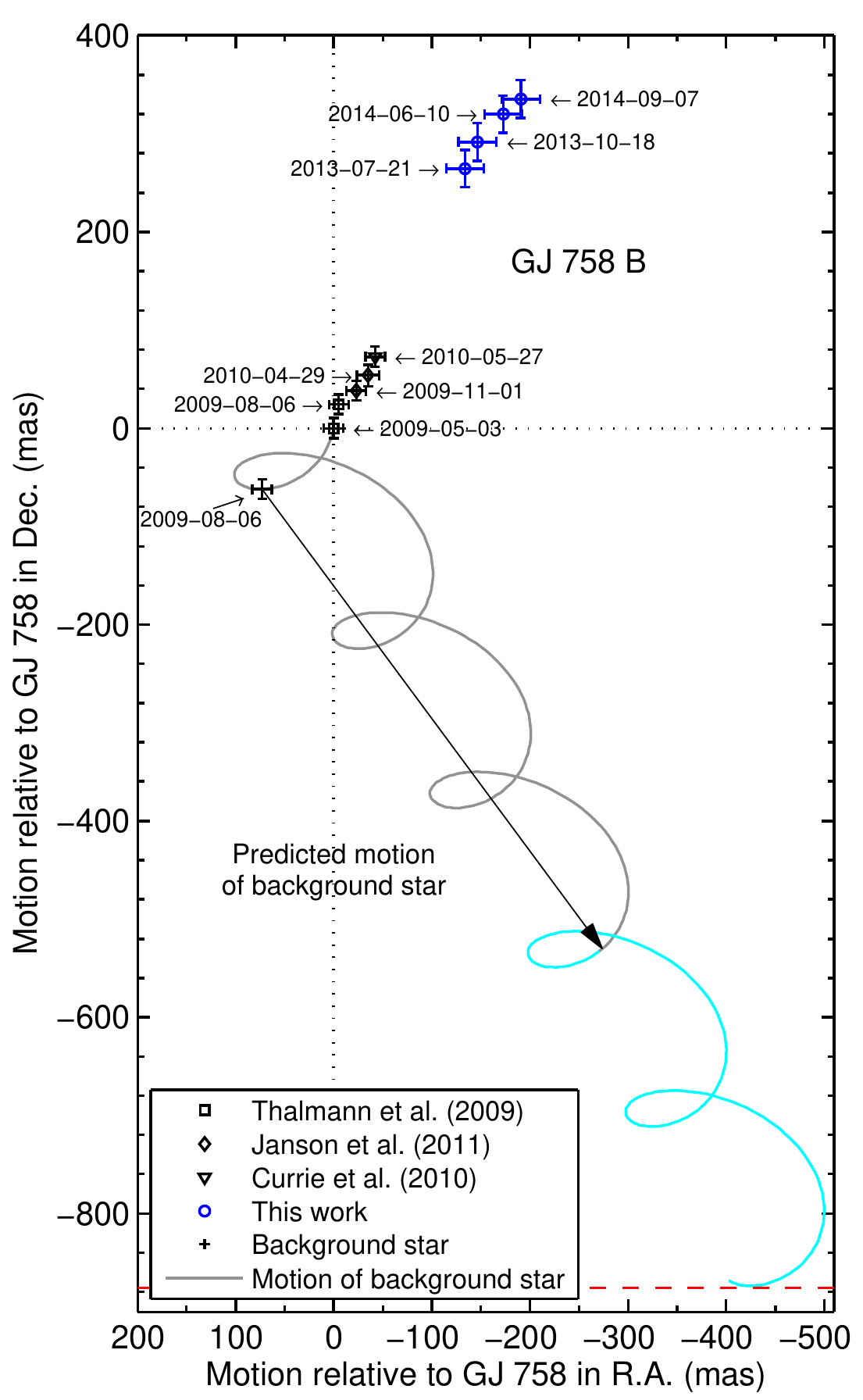}
\caption{Astrometric analysis of GJ\,758\,B, showing its position and implied northwest motion relative to the primary, since its discovery in 2009-05. The black markers are from previous astrometry, and the blue markers are from the four epochs of detection presented in this work, all with associated error bars displayed. The predicted motion (due to GJ\,758's proper motion and annual parallactic motion) of the background star discovered in 2009-08 is shown as a black curve, turning cyan at our first observing epoch in 2012-06. The star was in our FOV, nearing the edge of our detector (marked with a red dashed line) for the last two epochs, but too faint to be detected.}
\label{fig:astrometry}
\end{figure}

From the processing of image cubes with CACS (Section~\ref{sec:datared:cacs}) we expect the primary star to be aligned and centered in the images to a precision of $\sim$\,0.2 pixels. A larger error on the astrometry comes from the isotropic Gaussian fit to the GJ\,758\,B peaks, which can be distorted by residual speckle noise. As a conservative precaution we present astrometric errors equivalent to $\pm$1\,pix in Fig.~\ref{fig:astrometry} and Table~\ref{tab:astrometry}.

The measured angular separation and PA of GJ\,758\,B relative to the primary star is presented in Table~\ref{tab:astrometry}, together with previous astrometry data. Fig.~\ref{fig:astrometry} shows the motion of GJ\,758\,B \changed{with respect to the primary} since the discovery by \citet{Thalmann2009} in 2009-05, and also includes the expected motion of a nearby faint object previously detected in three epochs \citep{Thalmann2009,Janson2011b} and determined to be a background star, but undetected in our observations. Because of GJ\,758's high proper motion \citep[${\mu}_{\mathrm{R.A.}} = 83.40\,\text{mas}\,\text{yr}^{-1}$ and $\mu_{\mathrm{Dec.}} = 162.32\,\text{mas}\,\text{yr}^{-1}$;][]{VanLeeuwen2007}, the background star should have moved more than 600\,mas to the southeast since its discovery in 2009-08 to our first observed epoch in 2012-06. In the time span of our five observed epochs, it would be expected to lie on the cyan colored curve in Fig.~\ref{fig:astrometry}, within the P1640 FOV with GJ\,758 centered on the detector, but very close to our southern edge (marked by a dashed red line) for our two last epochs.

GJ\,758\,B is moving northwest \changed{as seen in the rest frame of GJ\,758\,A}, but so far with no discernible orbital curvature in nine epochs of astrometric data spanning almost 5.5\,years. A linear extrapolation from previous astrometry gave a good prediction on the final measured positions of the companion. A forthcoming paper will discuss orbital simulations (J.~Aguilar et al.\ 2017, in preparation).

The projected separation of the companion to the GJ\,758 primary is $\sim$26\,au in our last (2014-09) epoch, using the parallax from the most recent Hipparcos reduction \citep{VanLeeuwen2007}.

\begin{deluxetable*}{ccccl}
\tabletypesize{\scriptsize}
\tablecaption{Astrometry for GJ\,758\,B}
\tablewidth{1.0\textwidth}
\tablehead{\colhead{Besselian Date} & \colhead{Julian Date} & \colhead{Position Angle, $\theta_\text{PA}$} & \colhead{Angular Separation, $\rho$} & \colhead{Reference} \\
						\colhead{(years)} & \colhead{(days)} & \colhead{($\degr$)} & \colhead{($\arcsec$)} & } 
\startdata
2009.338 & 2454955.2 & $197.77\pm0.15$ & $1.879\pm0.010$ & \citet{Thalmann2009}  \\
2009.598 & 2455050.1 & $198.18\pm0.15$ & $1.858\pm0.010$ & \citet{Thalmann2009}  \\
2009.836 & 2455137.1 & $198.83\pm0.31$ & $1.850\pm0.010$ & \citet{Janson2011b}  \\
2010.326 & 2455316.0 & $199.34\pm0.34$ & $1.839\pm0.011$ & \citet{Janson2011b}  \\
2010.403 & 2455344.1 & $199.76\pm0.15$ & $1.823\pm0.015$ & \citet{Currie2010}  \\
2012.461 & 2456095.9 & $204.92\pm0.31$ & $1.680\pm0.019$ & This work  \\
2013.553 & 2456494.8 & $205.69\pm0.31$ & $1.661\pm0.019$ & This work  \\
2014.441 & 2456818.9 & $206.95\pm0.29$ & $1.648\pm0.019$ & This work  \\
2014.684 & 2456907.7 & $207.75\pm0.28$ & $1.643\pm0.019$ & This work 
\enddata
\tablenotetext{}{}
\label{tab:astrometry}
\end{deluxetable*}

\subsection{Optimization of S4 speckle-suppression parameters}\label{sec:pc_opt}

\begin{figure*}[htb]
\center
\resizebox{1.0\hsize}{!}{\includegraphics[width=.45\linewidth]{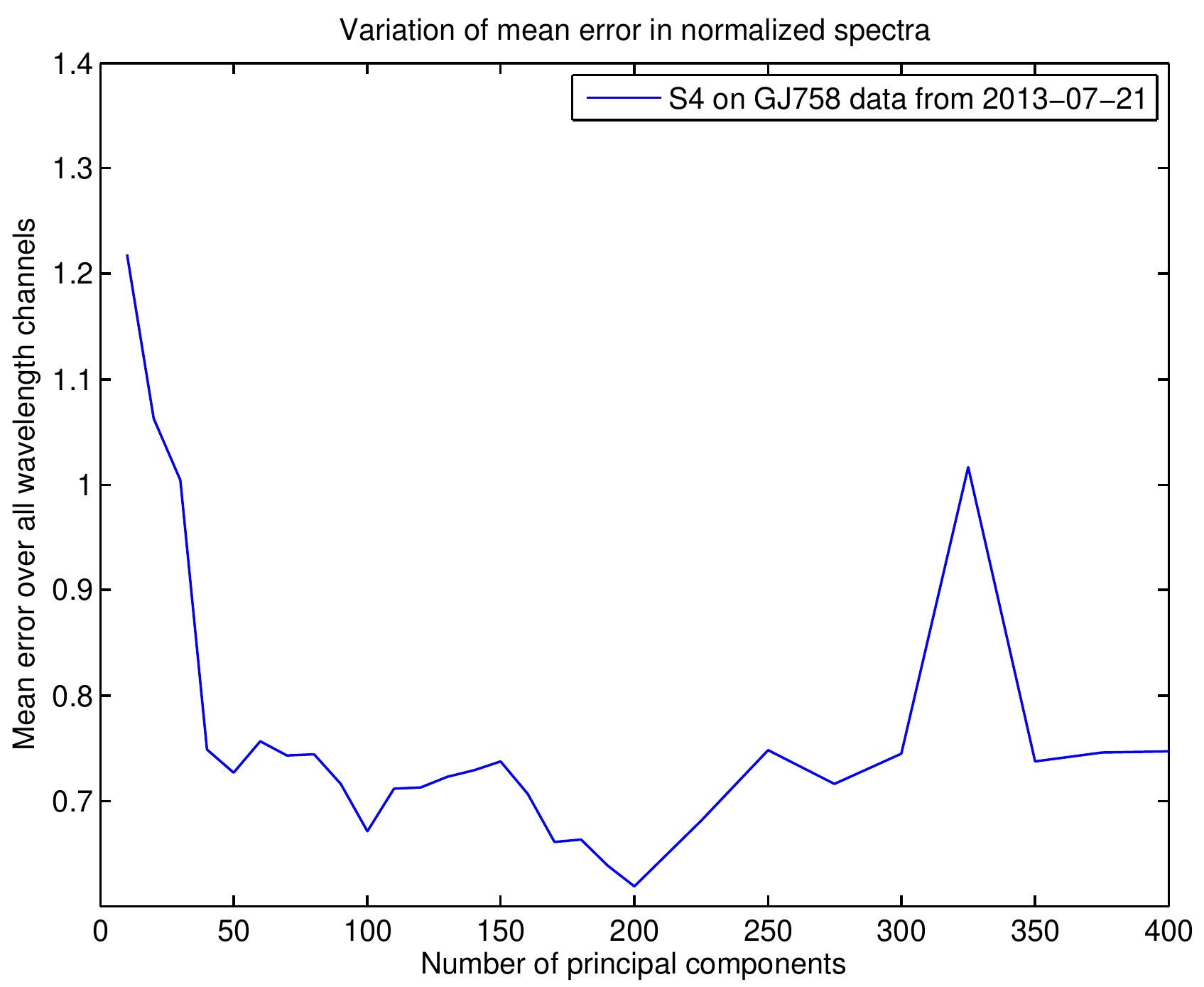} \includegraphics[width=.55\linewidth]{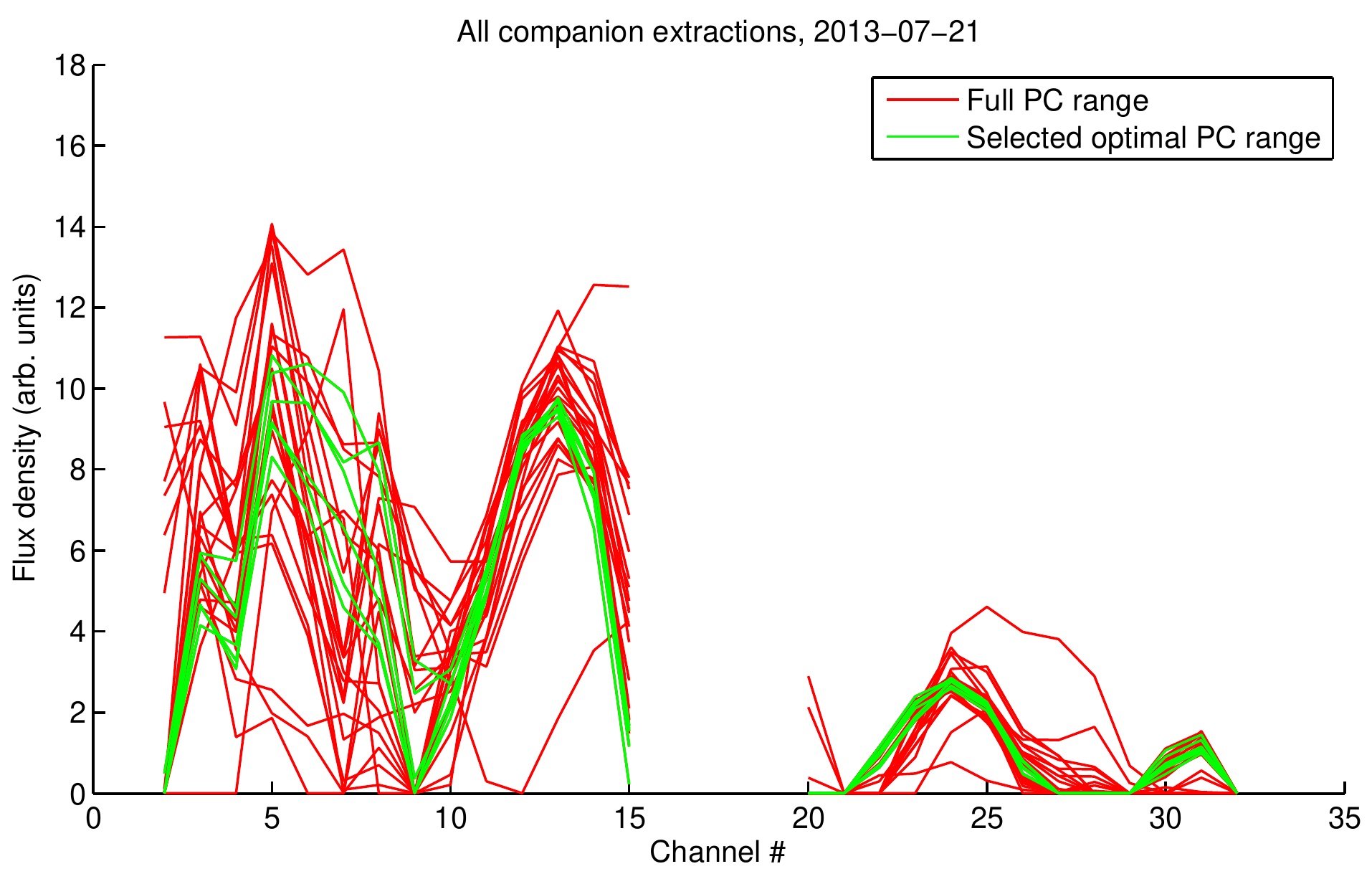}}
\caption{{\emph{Left:} plot of the mean noise over all wavelength channels for the S4 spectral extraction of GJ\,758\,B in 2013-07, over the full range of principal components examined. An overall minimum is generally reached for $N_\text{PC} \approx 100$--250. \emph{Right:} extracted companion spectra from the 2013-07 epoch for the full $N_{\mathrm{PC}}=\{10,20,30,...,400\}$ sample in each wavelength channel plotted in red, with the selected optimized ($N_{\mathrm{PC}}=$ \{160, 170, 180, 190, 200, 225\}) sub-sample plotted in green.}}
\label{fig:s4s_optim}
\end{figure*}

For spectral extractions, the width of the test region in polar coordinates is $\Delta\theta = 5$ pixels (as opposed to the $\Delta\theta = 3$ pixels used in the S4 detection phase). This has in previous tests of S4 given a better signal extraction than $\Delta\theta = 3$ or 7. Using the 1-$\sigma$ errors derived from fake insertions as described in Section~\ref{sec:datared:s4s}, we calculate the mean error over all wavelength channels in the normalized spectra, and plot them versus the number of PCs, $N_\text{PC}$, used in the PCA image reconstruction and spectral extraction. As can be seen in the left panel of Fig.~\ref{fig:s4s_optim}, a minimum is reached at $N_\text{PC} \approx 160$--225 for the 2013-07 epoch. {The typical range of optimal $N_\text{PC}$ found in S4 spectral extraction is consistently between 100 and 250, depending on data quality. For the final spectra of each epoch, we calculated error-weighted mean spectra from $N_{\mathrm{PC}}=$ \{160, 170, 180, 190, 200, 225\}, \{190, 200, 225, 250\}, and \{120, 130, 140, 150, 160, 170, 180, 190\}, for 2013-07, 2014-06, and 2014-09, respectively. For 2013-10 we did not obtain a clear minimum and decided to include the full $N_\text{PC} = 10$--400 range. }

To show the total spread in extracted spectra over all tested number of PCs, $N_{\mathrm{PC}}=\{10, 20, 30,..., 400\}$, we plot them together with the $N_{\mathrm{PC}}$ optimized sub-sample {for the 2013-07 extraction in the right panel of Fig.~\ref{fig:s4s_optim}.}

\subsection{Final GJ\,758\,B spectrum} \label{sec:res:spectrum}
\begin{figure}[htb]
\includegraphics[angle=0,width=0.5\textwidth]{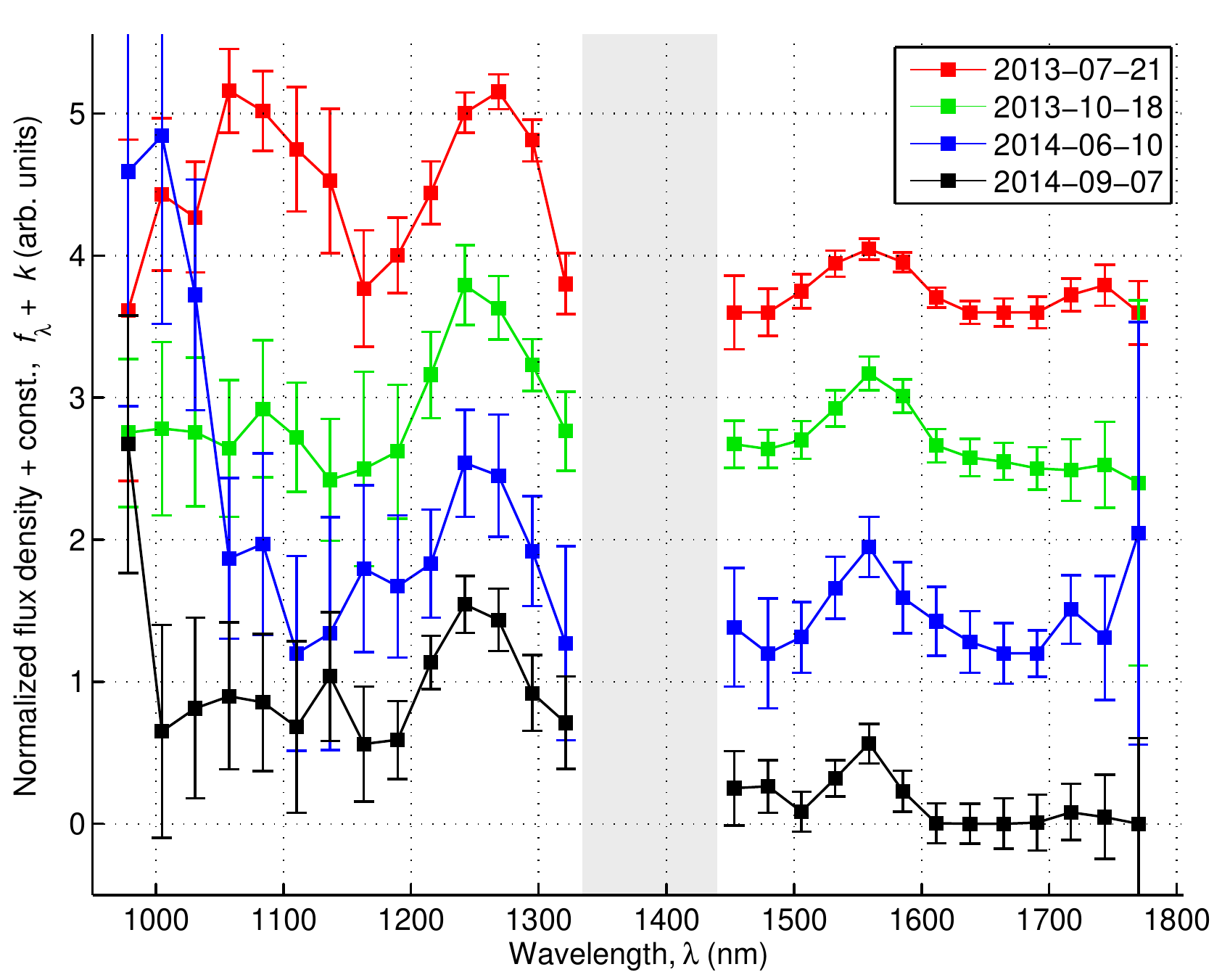}
\caption{S4 extracted spectra of GJ\,758\,B from all four epochs in which it was detected, labeled with observing date as YYYY-MM-DD. Error bars show 1-$\sigma$ spectral deviations of 50 fake white-light sources inserted and extracted at the same projected angular distance from the primary star as the companion, averaged over the included $N_{\mathrm{PC}}$ range found from noise optimization (see Section~\ref{sec:pc_opt}).}
\label{fig:GJ758_spectrum_obs}
\end{figure}

Extracted final spectra from the four detected epochs are overplotted in Fig.~\ref{fig:GJ758_spectrum_obs}. Error bars represent 1-$\sigma$ deviations derived from fake source insertions and extractions, as explained in Section~\ref{sec:datared:s4s}, using the weighted error over included optimal $N_{\mathrm{PC}}$ range (and $\Delta\theta=5$\,pixels). {As an example of} the variation between extracted spectra for the explored $N_{\mathrm{PC}}$ range{, all extractions for the 2013-10 epoch are presented in Section~\ref{sec:pc_opt}.} Parameter optimization for processing of the data from 2014-06 was difficult due to its bad quality (severe ``mirror seeing''), thus the spectrum from that date has lower S/N, and may be considered only a marginal detection. Maximum S/Ns are listed as detection significance in Table~\ref{tab:obs}. All spectra have been normalized by division with their mean flux density in the observed wavelength range. Channels in the photometric $Y$ band, covering 960--1080\,nm, are very noisy due to our decreased sensitivity toward shorter wavelengths, with no significant detection of the companion in either epoch. The same holds true over 1350--1480\,nm, the shaded region in Fig.~\ref{fig:GJ758_spectrum_obs}, where we have bad sensitivity due to absorption from telluric atmospheric water vapor. That is also a region where have large uncertainties in the applied SRF since the spectral calibrators from which it is derived were observed at different times and thus in different conditions. The most striking features in all spectra are sharp peaks around 1280 and 1580\,nm --- a spectral signature characteristic of T~dwarfs (see Section~\ref{sec:ana:atm}). Overall, the agreement between spectra in the $J$ (1014--1327\,nm) and $H$ (1477--1784\,nm) bands is good (within 2$\sigma$), and we use the combined spectrum (weighted average with weighted errors) for further modeling and analysis in the next section.
% Rephrased sentence about non-detection in Y

\subsubsection{KLIP confirmation of S4 spectral extraction}\label{app:s4_vs_klip}
\begin{figure}[h]
\includegraphics[width=0.5\textwidth]{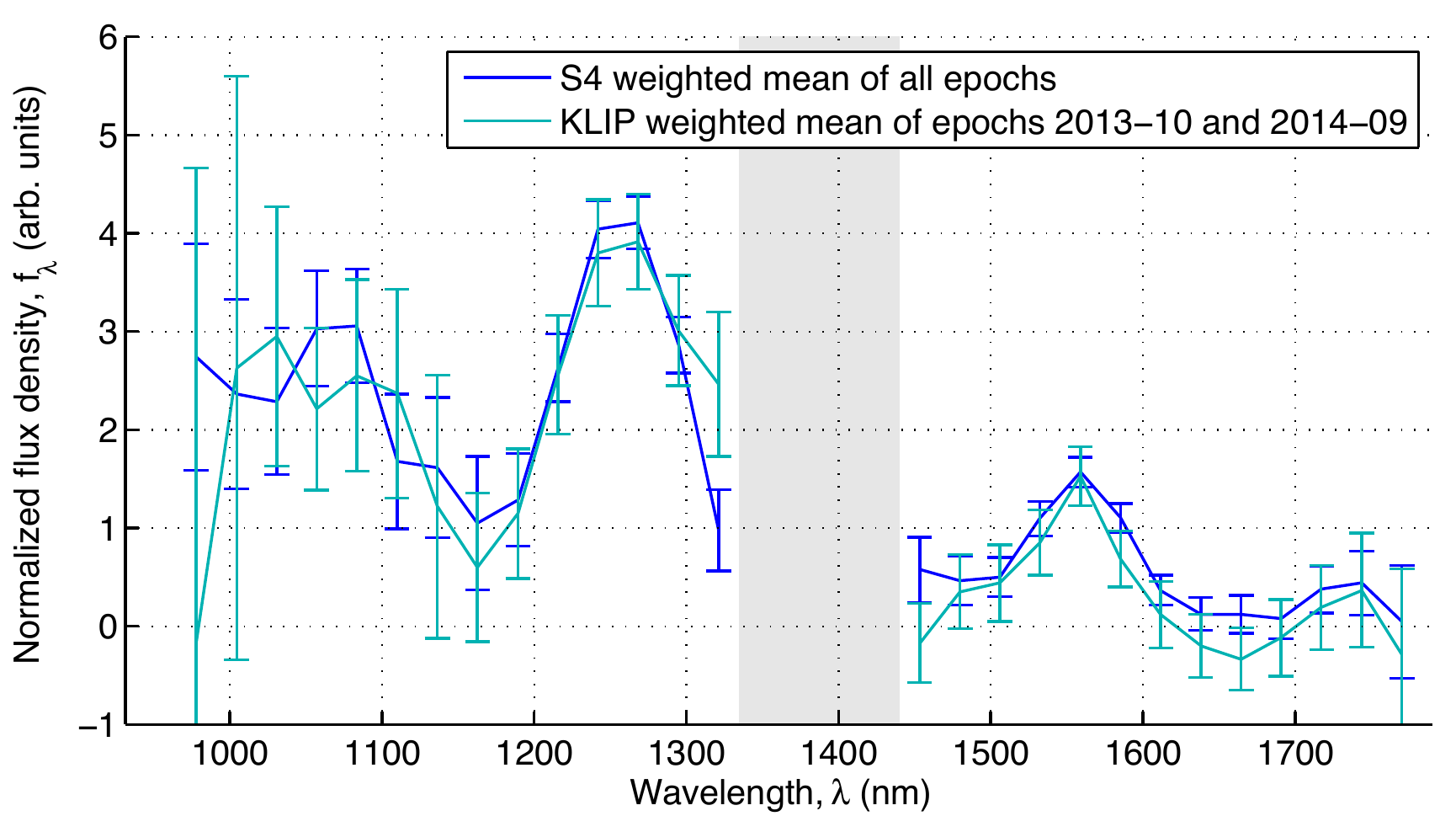}
\caption{The KLIP extracted spectrum of GJ\,758\,B combined from the 2013-10-18 and 2014-09-07 observations, compared to the S4 extracted spectrum combined from 2013-07-21, 2013-10-18, 2014-06-10, and 2014-09-07 data.}
\label{fig:s4allepochs_vs_kliptwoepochs}
\end{figure}

In addition to extracting the P1640 observed spectrum of GJ\,758\,B using S4, we independently confirmed the extraction using KLIP. The companion was immediately detected in KLIP reductions of observations on 2013-10-18 and 2014-09-07. Fig.~\ref{fig:s4allepochs_vs_kliptwoepochs} shows how well the combined KLIP extractions from those two epochs match the combined S4 extractions from all four detected epochs. Additional modeling of the combined KLIP extracted spectra following the procedure in Section~\ref{sec:ana:atm} confirms the inferred spectral type, effective temperature and surface gravity of the companion to within stated errors.

\begin{deluxetable*}{lccccc} % Date | Julian date | S/N_Y | S/N_J | S/N_H | S/N_lambda,max. \\ Combined S/N
\tabletypesize{\scriptsize}
\tablecaption{{S/N of GJ\,758\,B detection in \emph{YJH} after locally optimized spectral extraction}}
\tablewidth{0pt}
\tablehead{\colhead{Date} & \colhead{Julian date} & \colhead{$\text{S/N}_{Y}$} & \colhead{$\text{S/N}_{J}$} & \colhead{$\text{S/N}_{H}$} & \colhead{$\text{S/N}_{\lambda,\text{max}}$\tablenotemark{a}} \\ \colhead{(UT)} & \colhead{(days)} & \colhead{($\sigma$)} & \colhead{($\sigma$)} & \colhead{($\sigma$)} & \colhead{($\sigma$)}}
\startdata
2013-07-21 & 2456494.8199421302 & 4.2 & 9.5 & 3.9 & 12.7 \\
2013-10-18 & 2456583.5734722228 & 1.7 & 4.7 & 3.0 & 6.4 \\
2014-06-10 & 2456818.9438657411 & 4.2 & 3.0 & 1.8 & 3.6 \\
2014-09-07 & 2456907.6952893524 & 4.5 & 7.7 & 3.2 & 7.7 \\
\tableline \\[-4pt]
\multicolumn{2}{c}{Combined SRN for all epochs} & 7.9 & 14.1 & 7.3 & -- 
\enddata
\tablenotetext{a}{Highest statistical significance of signal in any wavelength channel from locally optimized spectral extractions.}
\label{tab:detection_levels}
\end{deluxetable*}

\section{Discussion} \label{sec:analysis}

This section contains further analysis of our results, presenting photometry, atmospheric modeling, and a prediction of GJ\,758's RV trend.

\subsection{Photometry}

P1640's simultaneous coverage of the $Y$, $J$, and $H$ bands, gives it an advantage over many other high-contrast imaging instruments when it comes to relative flux calibration and the object properties derived from near-IR photometric colors. Assuming a well-characterized SRF, it essentially obtains automatic relative flux over all channels, without adding uncertainties from calibration of separate observations taken with different filters. Post-processing introduces additional uncertainties in the absolute flux, which can be quantified using fake insertions or $N_\text{PC}$ plateaus. Despite this challenge, we can nevertheless estimate the $J-H$ color of GJ\,758\,B. After multiplying with the Gemini/\changed{NIRI MKO} filter transmission curves used in \citet{Janson2011b}, we integrate the normalized flux densities over $J$ and $H$ to obtain \changed{$J-H = -0.7 \pm 0.3$\,mag}. This is consistent (within the errors) with the $J-H = -0.58 \pm 0.28$\,mag derived by \citet{Janson2011b}\changed{, confirming its exceptionally blue near-IR color, similar to cool cloud-free field T~dwarfs \citep{Kirkpatrick2011,Dupuy2012}. The small change in $J-H$ color along the emperical T~dwarf sequence for field objects makes it hard to determine a more exact spectral type based on color alone.} Our atmospheric analysis below does however indicate that GJ\,758\,B could possibly be a slightly warmer and earlier type BD than previously found.\footnote{This is also consistent, depending on interpretation, with the SPHERE results presented by \citet{Vigan2016} during the referee processes of this paper.}

\subsection{Spectral Fitting} \label{sec:ana:atm}
Below we use both observed and synthetic spectra of T~dwarfs to perform spectral typing and derive physical characteristics of GJ\,758\,B \changed{based on the spectrum obtained with P1640. It is pertinent here to note that deriving any astrophysical object's effective temperature and surface gravity is inherently dependent on the model used and the data included, be it broad- or narrow-band photometry, low-resolution spectra, high-resolution spectra, different wavelength ranges, etc. The spectrophotometric fitting used here can give different answers than broader SED fitting, both of which can give different answers from atomic line ratios -- all from the same model. In this sense, ``temperature'' is a somewhat philosophical value, with different methods and models producing systematically different numbers for the same object. Some authors have, for example, plotted spectra of these sorts of objects as brightness temperature versus wavelength to illustrate how each part of the spectrum arises from very different depths in the object's atmosphere \citep{Matthews1996,Oppenheimer1999}.}

\subsubsection{Empirical analysis}
An effective way to estimate the spectral type of T~dwarfs is by comparison of their near-IR (0.8--2.5\,$\mu$m) spectra to those of standard T~dwarfs \citep{Burgasser2006}. Here, we compare the combined \emph{YJH} GJ\,758\,B spectrum to 154 T~dwarfs, ranging from spectral types T0 to T9, including the T0--T8 standards, in order to estimate its spectral type. Most of the spectra were obtained from the SpeX Prism Library.\footnote{\texttt{http://pono.ucsd.edu/\~{}adam/browndwarfs/spexprism/}}
% Changed subsubsection heading 
 
\begin{figure}
\includegraphics[trim=0.3cm 0.5cm 0.51cm 1.5cm, clip=true, width=0.5\textwidth]{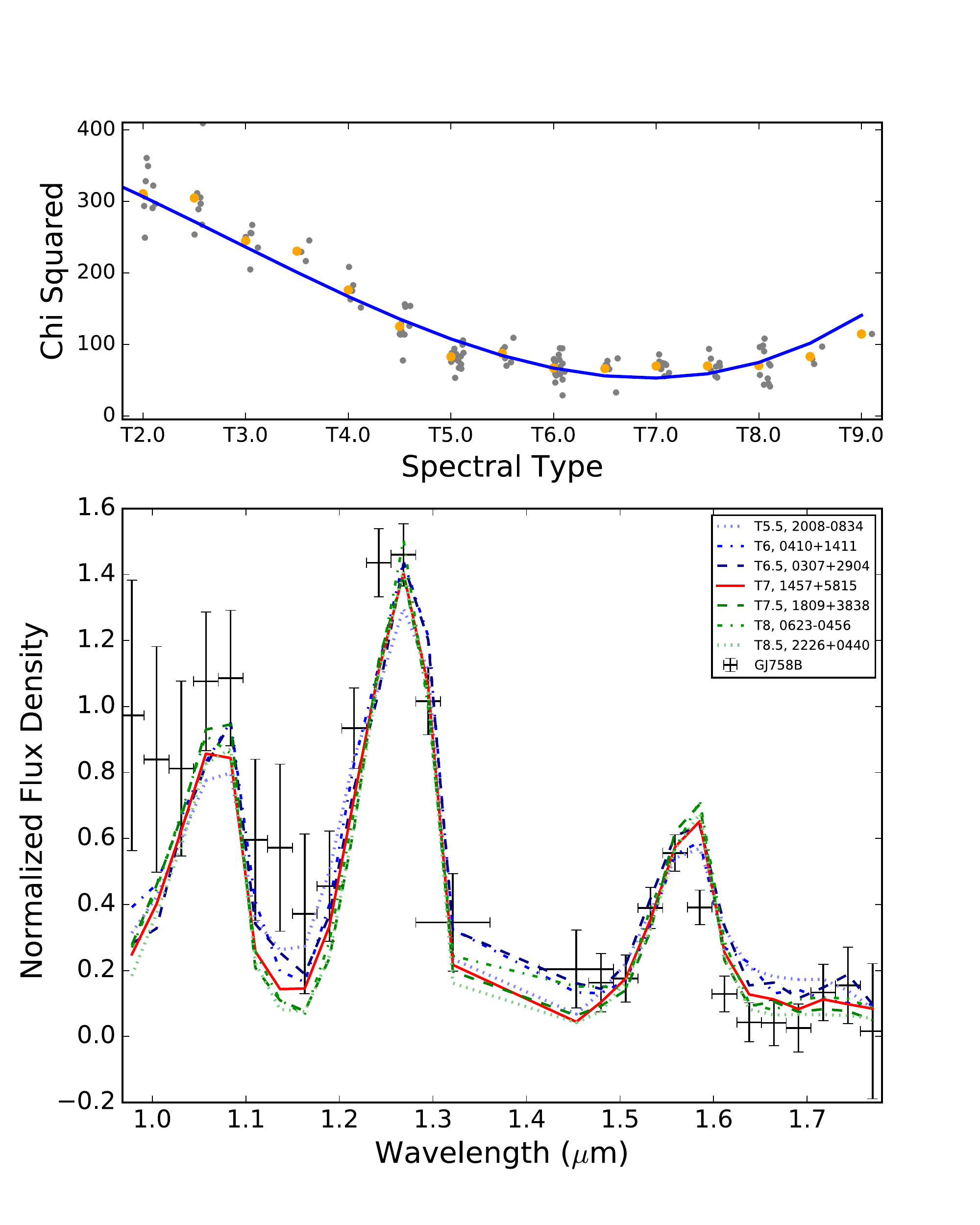}
\caption{(Top) $\chi^2$ as a function of spectral type for T0--T9 objects. {A third-order (blue) polynomial fit from T0--T9 based on the average $\chi^2$ results per spectral type, and the average $\chi^2$ points per spectral type (orange), are also shown. We derive a spectral type of $\text{T}7.0\pm1$} from this spectral comparison. (Bottom) P1640 spectrum of GJ\,758\,B (black crosses) plotted with binned and trimmed SpeX Prism spectra of T5.5--T8.5 objects. }
\label{fig:TK}
\end{figure}

The SpeX Prism spectra are binned to the resolution of the P1640 observation by adding the flux (and uncertainties in quadrature) within the wavelength bins appropriate to the lower resolution spectrum. A $\chi^2$ is calculated for each binned T~dwarf spectrum compared to the GJ\,758\,B spectrum. We have removed the water band around 1.4 $\mu$m from the fit, though fits were completed both with and without those flux points and the results were essentially identical. Errors from both the GJ\,578\,B spectrum and the binned T~dwarf templates are used in the calculation. Fig.~\ref{fig:TK} (top) shows $\chi^2$ as a function of spectral type for each T~dwarf template with {a third order polynomial fit. Using the S4 extracted spectrum, a spectral type of T7.0 \citep[WISE\,J145715.03+581510.2,][]{Kirkpatrick2011} results in the minimum $\chi^2$ value of $\sim$60 with 29 degrees of freedom.\footnote{Note that our $\chi^2$ vs.\ spectral type relation looks almost identical to the goodness-of-fit value vs.\ spectral type plot in Fig.~5 of \citet{Vigan2016}, also with a minimum of T7.0 after visual inspection}. Fig.~\ref{fig:TK} (bottom) shows this best fit, along with example fits, also of lowest $\chi^2$, for the surrounding T~dwarf spectral subtypes (WISE\,J200804.71−083428.5, WISE\,J180901.07+383805.4, \citet{Mace2013}; WISE J041054.48+141131.1, WISE J030724.59+290447.4, WISE J062309.94−045624.6, WISE J222623.05+044004.0, \citet{Kirkpatrick2011}). Based on these comparisons we estimate a spectral type of $\text{T}7.0\pm1$ for GJ\,758\,B.}

\subsubsection{Model Atmospheres}

T~dwarfs are classified according to their near-IR spectra, but spectral types do not necessarily correspond directly or uniquely to physical properties \citep[e.g.,][]{Kirkpatrick2008}. To constrain atmospheric parameters, in this case effective temperature and surface gravity, we compare the observed P1640 spectrum of GJ\,758\,B to synthetic spectra from the BT-Settl13 model atmospheres \citep{Allard2014}. We use a grid of solar metallicity models with effective temperatures from 400--4500\,K in increments of 50 or 100~K and surface gravities of $\log g = 3.5,~4.0,~4.5,~5.0$, and 5.5 (cgs units). The model fitting procedure is based on that described in \citet{Rice2015}. Versions of the fitting procedure have been applied to P1640 spectra in \citet{Roberts2012a}, \citet{Hinkley2013}, and \citet{Crepp2015}. The fitting procedure is summarized briefly below.

Model spectra are binned from their native resolution of $\Delta\lambda = 0.1$\,nm to match that of the P1640 {($\Delta\lambda = 26.4$\,nm)} spectrum. A goodness-of-fit parameter similar to $\chi^2$ \citep[see][]{Cushing2008} is calculated for each fit of the observed spectrum to each binned spectrum in the model grid. The model parameters for the spectrum with the minimum $\chi^2$ are used as the starting point for generating probability distributions, $P \propto \exp −\chi^2/2$, using a $10^6$-step Markov Chain Monte Carlo (MCMC) analysis using the Metropolis Hastings algorithm. The MCMC routine interpolates between calculated model spectra as it moves along the chain of steps. We find that jump sizes of 200\,K in temperature and 2.0\,dex in surface gravity provide optimal acceptance ratios of $\sim$\,0.3--0.4.

\begin{figure}
\includegraphics[trim=2.5cm 1.0cm 1.0cm 1.5cm, clip=true, width=0.5\textwidth]{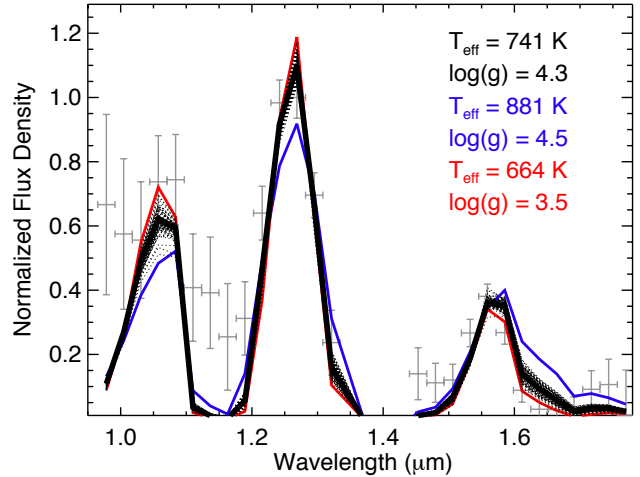}
\caption{{Best fit synthetic spectrum from BT-Settl13 models (black) using parameters derived from MCMC analysis of the complete \emph{YJH} spectrum (excluding the terrestrial water absorption band) from S4 extractions of the P1640 spectrum of GJ\,758\,B (gray markers and error bars). 100 spectra were randomly chosen from the posterior distributions of the MCMC calculations to represent the range of model fits that are allowed within $1-\sigma$ uncertainty for the spectrum. The best fit parameters, 741\,K/4.3\,dex~(cgs) for the \emph{YJH} spectrum, are the 50\% quantiles of the effective temperature and surface gravity parameters from the MCMC posterior distributions (see Fig.~\ref{fig:post_dist}. The BT-Settl13 spectra with parameters determined from \emph{YJ} and $H$ fits only are shown in red and blue, respectively.}}
\label{fig:synth_spec_fit}
\end{figure}

{We fit model spectra to three versions of each extracted spectrum of GJ\,758\,B (and 100 random samples from the MCMC chain): (1) the complete (\emph{YJH}) spectrum without the four flux points closest to the water absorption band at 1.4\,$\mu$m, (2) the flux points blueward of 1.33\,$\mu$m (\emph{YJ}), and the flux points redward of 1.45\,$\mu$m ($H$). Results for all three versions of the spectra from each extraction are presented in Figure~\ref{fig:synth_spec_fit} in black, red, and blue, respectively.}

\begin{figure}
\includegraphics[trim=0cm 0cm 0cm 0cm, clip=true, width=0.5\textwidth]{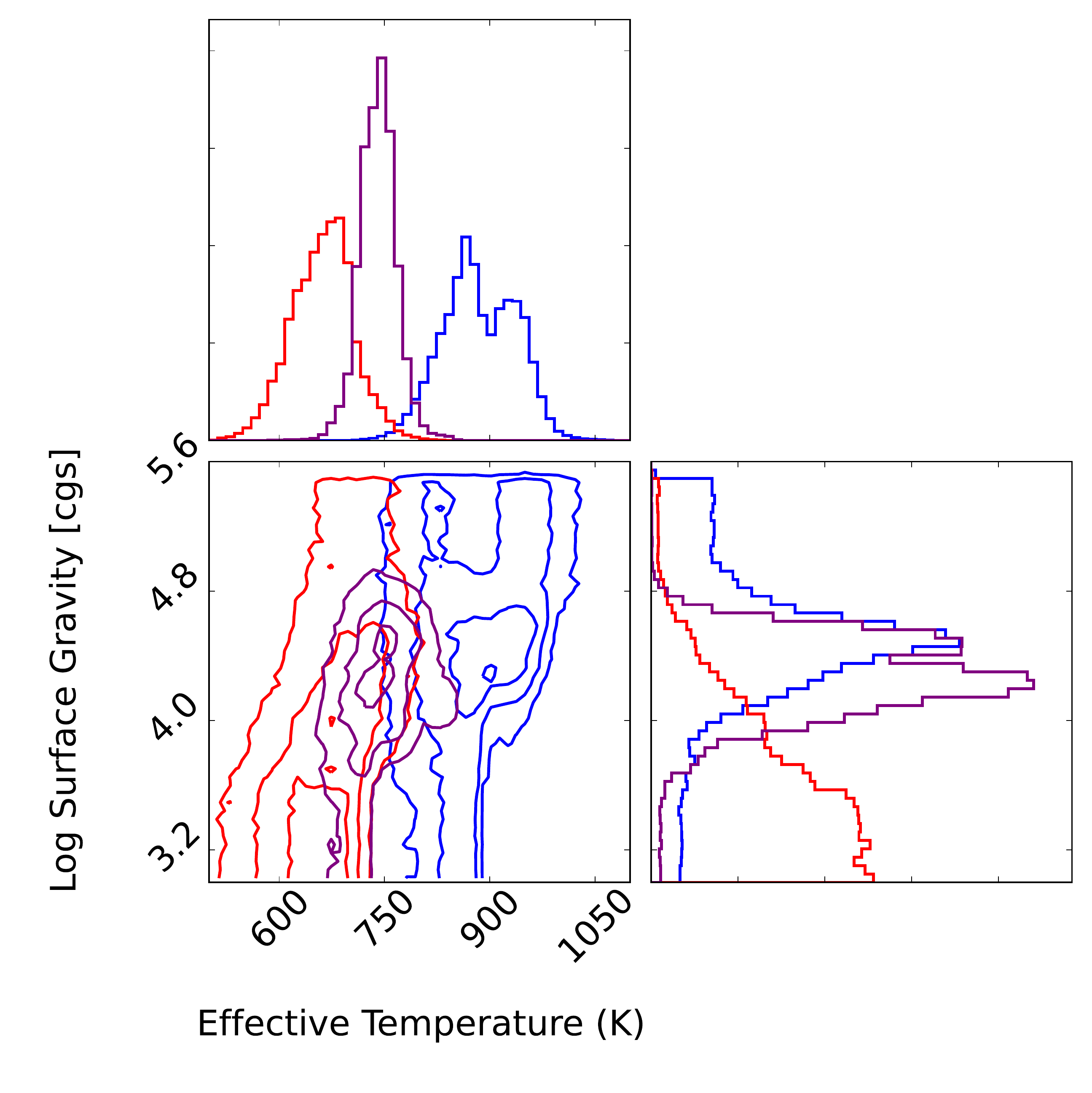}
\caption{Posterior distributions of the MCMC analysis using $10^6$ steps for the complete \emph{YJH} spectrum {(purple), $YJ$ (red), and $H$ (blue)} S4 extracted spectrum of GJ\,758\,B. Histograms show the distributions marginalized over gravity (top left) and temperature (bottom right). Model fits to the low-resolution near-infrared spectrum from P1640 provide a better constraint in temperature than in surface gravity. GJ\,758\,B has a temperature of {$T_\mathrm{eff} = 741 \pm 25$\,K} and a surface gravity of {$\log g = 4.3 \pm 1.0$\,dex~(cgs)} (1-$\sigma$ uncertainties).}
\label{fig:post_dist}
\end{figure}

{Posterior distributions from the subsequent MCMC analysis, presented in Figure~\ref{fig:post_dist}, show a range in $T_\mathrm{eff}$ from $\sim$620 to $\sim$940\,K for the three spectra. The posterior distributions marginalized over $\log{g}$ show a clear peak in $T_\mathrm{eff}$ for the complete spectrum, with broader and slightly asymmetric histograms for \emph{YJ} and $H$ fits. The $\log{g}$ histograms cover a wide parameter range for the three fits, with the peak for $H$ fit falling below  the edge of the model grid at low surface gravities. The peak at higher surface gravity values is higher and corresponds to slightly hotter $T_\mathrm{eff}$ values. As was noted in \citet{Crepp2015} and \citet{Rice2015}, we currently cannot reliably infer gravity from these low-resolution spectra. However, the higher end of the range is consistent with the gravity predicted by evolutionary models \citep[e.g.,][]{Baraffe2003} based on the temperature of GJ\,758\,B and the age constraint provided by the primary star. }

Using the mode posterior distributions and $68\%$ confidence interval for the posterior of the spectral fits, we find that GJ\,758\,B has {best-fit $T_\textrm{eff}$ and $\log{g}$ from the MCMC of $741\pm25$\,K and $4.3\pm0.5$\,dex for \emph{YJH}, $881\pm60$\,K and $4.5\pm0.5$ for $YJ$, and $664\pm45$ and $3.5$ for $H$. The value for the complete \emph{YJH} spectrum is consistent within the uncertainty to temperatures predicted by the spectral type $T_\text{eff}$ relationships of \citet{Filippazzo2015} for a T7.0 object ($825 \pm 113$\,K for the M6­­--T9 relationship).} 

The $\text{T}7.0 \pm 1.0$ spectral type we find from the spectral analysis of GJ\,758\,B is slightly earlier than the prediction of T8--T9 based on near- and mid-infrared photometry \citep{Janson2011b}, as was the case for HD 19467\,B \citep[T5--T7 predicted from photometry, $\text{T}5.5 \pm 1.0$ from the P1640 spectral analysis][]{Crepp2015}. For the later spectral type and cooler temperature of GJ\,758\,B, the surface gravity results are more consistent with predictions from evolutionary models {($\log{g}=3.0$--5.0)}, but the broad MCMC posterior distributions show that model fits to low-resolution near-IR spectra are still unreliable for independent age confirmation \citep[see also][]{Rice2015}. Based on COND03 evolutionary models \citep{Baraffe2003} for the temperature of the companion and age of the primary star, we infer a mass of 40--50\,$M_\mathrm{Jup}$ for GJ\,758\,B. The derived properties of GJ\,758\,B are summarized in Table~\ref{tab:properties}.

\begin{deluxetable}{lc}[h]
\tabletypesize{\scriptsize}
\tablecaption{Derived properties for GJ\,758\,B}
\tablewidth{0.4\textwidth}
\tablehead{\colhead{Property} & \colhead{GJ\,758\,B}}
\startdata
Spectral type & {T7.0$\pm$1} \\
Effective temperature, $T_\text{eff}$ & {$741{\pm}25$\,K} \\
Surface gravity, $\log g$ & {$4.3{\pm}0.5$\,dex (cgs)} \\
Mass, $M$ (inferred) & 40--50\,$M_\text{Jup}$ 
\enddata
\tablenotetext{}{}
\label{tab:properties}
\end{deluxetable}

\begin{figure}
\includegraphics[trim=0cm 0cm 0cm 0cm, clip=true, width=0.5\textwidth]{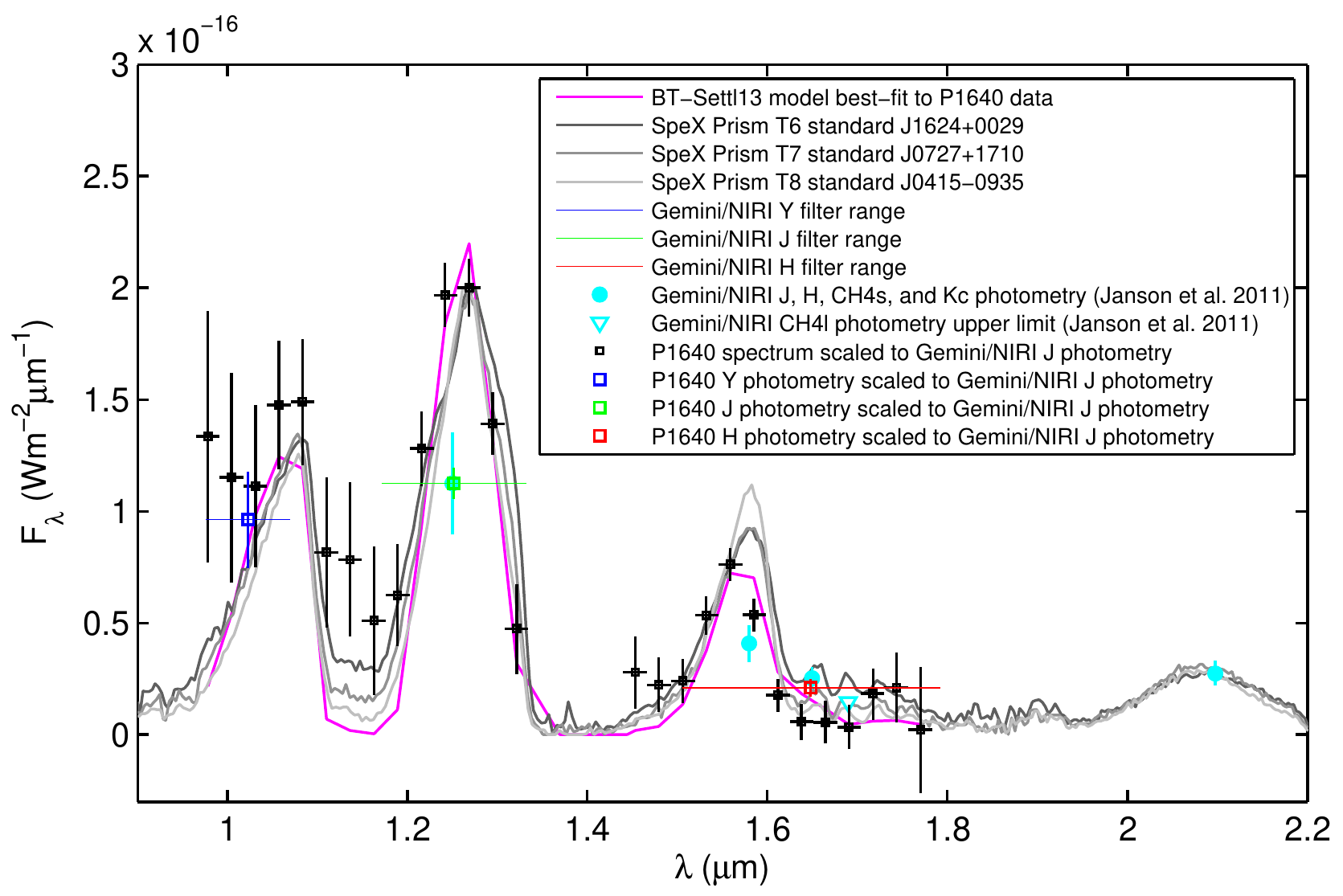}
\caption{\changed{The obtained P1640 spectrum and derived \emph{YJH} fluxes scaled by a factor found from matching Gemini/NIRI MKO $J$ band photometry with corresponding $J$ flux from P1640. Observed spectrum is shown in black, derived $YJH$ fluxes in blue, green, and red, respectively (with thin horizontal bars showing the filter bandwidth). Previous photometry from \citet{Janson2011b} is shown in cyan. The lines show the BT-Settl13 model and the SpeX Prism Library spectral type standards best-fit to the P1640 spectrum.}}
\label{fig:specphotadapt}
\end{figure}

\changed{Although not included in our modeling, due to the difficulty and large uncertainties involved in deriving absolute fluxes from this particular P1640 dataset, we plot previous photometry from \citet{Janson2011b} together with our data and spectral fits in Fig.~\ref{fig:specphotadapt}. The P1640 data have been scaled to the Gemini/NIRI $J$ band photometry, by calculating the corresponding P1640 flux from multiplication of the spectrum with the MKO $J$ filter transmission curve, integrating over the bandwidth, and matching the two flux densities. As can be seen in the figure, the resulting P1640 $H$ band flux also essentially overlaps with that of \citet{Janson2011b}. The spectrum of GJ\,758\,B is not well-matched by other empirical data of similar objects, specifically T6--T8 SpeX Prism standards, especially in the $H$ band peak near $1.58\,\mu$m. $K_{c}$ data at $2.1\,\mu$m does not seem to be a discriminator for spectral type, but such longer wavelength data \citep[including $L^{\prime}$ and \emph{Ms},][not shown in figure]{Janson2011b} will be important in establishing $\log g$, $T_\text{eff}$, [Fe/H], and cloud cover from current \citep[e.g.,][]{Morley2012,Saumon2012,Allard2014} and future improved atmospheric models.}

\subsubsection{Atmospheric composition}
\begin{figure}[h]
\includegraphics[width=0.5\textwidth]{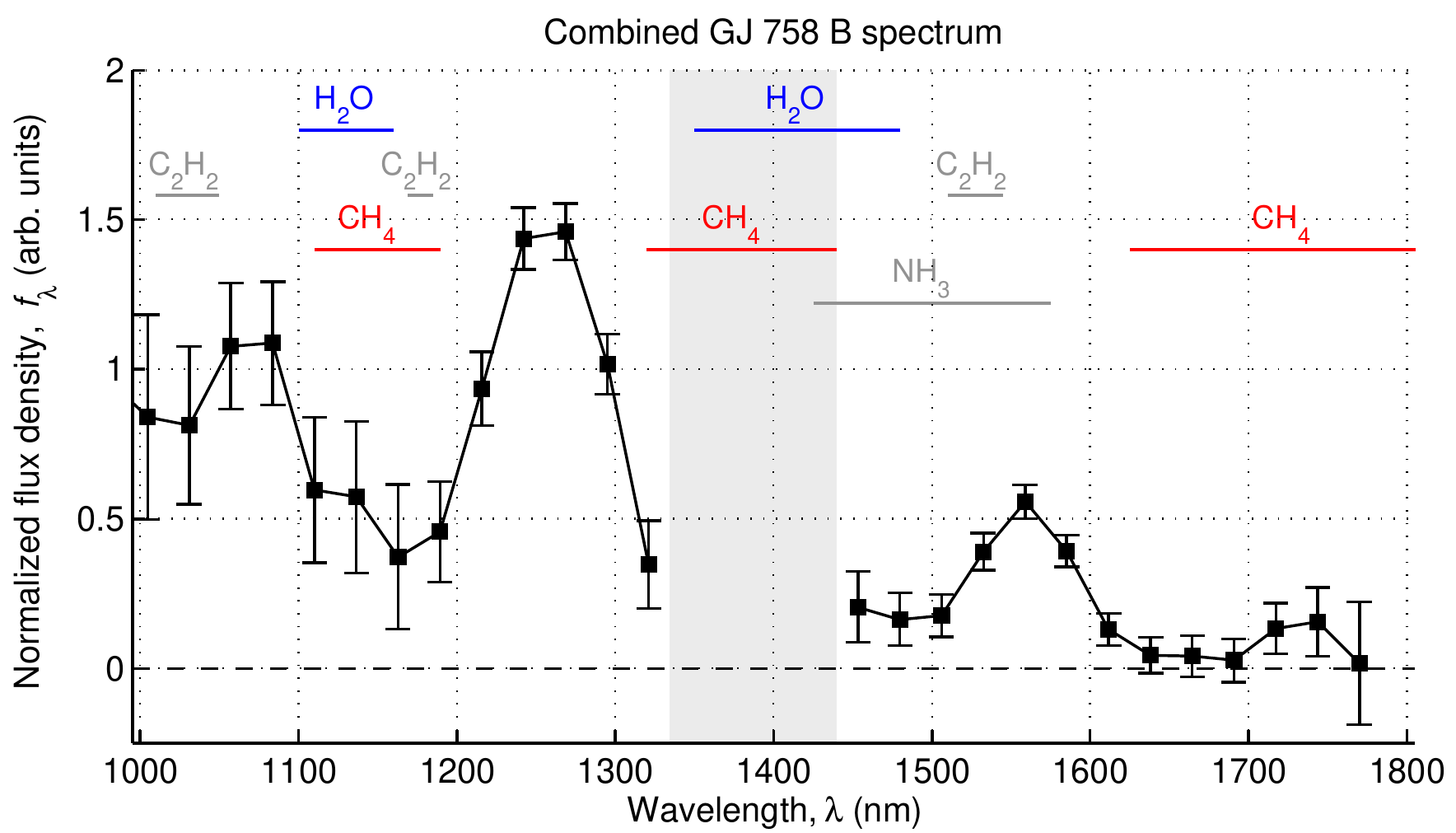}
\caption{A normalized weighted average spectrum of GJ\,758\,B with weighted error bars, shown together with the expected location of molecular absorption bands in this wavelength range. There is clearly strong absorption of CH$_4$ in the atmosphere, as expected for a cool T~dwarf, while no identification of other molecular species is implied.}
\label{fig:molec}
\end{figure}

One of the main drivers for obtaining spectra of substellar companions to stars is the prospect of not only determining global characteristics, like $T_{\mathrm{eff}}$, $\log{g}$, and $M_{*}$, but also say something about the chemical composition of their atmospheres. Multi-band photometry of GJ\,758\,B by \citet{Janson2011b} suggested clear methane (CH$_4$) absorption, which is to be expected in T~dwarfs, with ever stronger absorption for cooler objects. Our spectrum clearly confirms the presence of CH$_4$, with a deep absorption feature beyond 1600\,nm, and possibly absorption superimposed on H$_2$O features at shorter wavelengths in our range. No other molecular species are identified, but the location of the most prominent molecular absorption bands in the P1640 wavelength range (NH$_3$ and C$_2$H$_2$) are included for reference, and plotted together with the combined spectrum, calculated as the weighted average with mean errors, in Fig.~\ref{fig:molec}.

\subsection{Prediction of radial velocity (RV) trend}
We use our new mass estimate of GJ\,758\,B, together with the 68\% confidence bounds on the orbital parameters given in \citet{Janson2011b}, to make a rough estimate of the possibly observable RV trend of GJ\,758. From Kepler's laws we find that the star's velocity in orbit around the system's barycenter can be approximated by:
\begin{equation}
V_{*} \approx \frac{M}{M_{*}} \sqrt{\frac{G M_{*}}{a}}.
\end{equation}
For a companion mass $M$ in the range of 40--50\,$M_\mathrm{Jup}$, and orbital parameters (semimajor axis $a$, inclination $i$, and period $P$) from Table~2 in \citet{Janson2011b}, we get a maximum RV amplitude,
\begin{equation}
V_\text{obs} = V_{*} \sin{i},
\end{equation}
equal to 214\,$\text{m}\,\text{s}^{-1}$ for an orbital period of 170\,years, and 53\,$\text{m}\,\text{s}^{-1}$ for a period of 843\,years. Based on weighted median parameters we obtain 132\,$\text{m}\,\text{s}^{-1}$ for a period of 299\,yr. This implies a change in observed RV of GJ\,758 with between 2.52 and 0.13\,$\text{m}\,\text{s}^{-1}$ per year for the 68\% confidence interval, or 0.88\,$\text{m}\,\text{s}^{-1}$ per year for median parameter values. These estimates will be further constrained with new orbital modeling using all nine astrometry data points in J.~Aguilar et al.\ (2017, in preparation), but the above back-of-the-envelope calculation suggest that an RV trend could potentially be observable within a decade of continuous monitoring. However, to obtain model independent dynamical masses from orbit analysis with high accuracy we need both RV and astrometry with measurable curvature for a good part of the orbit. At that point, \object{GJ\,758\,B} could join \object{HD\,19467} \citep{Crepp2014} as one of the coolest sub-stellar objects with dynamical mass and spectrum, making it an important benchmark object for brown dwarf studies.

\section{Conclusions}
We have detected GJ\,758\,B in four epochs, and obtained the first near-IR spectrum of this substellar companion to a Sun-like star. Based on atmospheric modeling we conclude the following:

\begin{enumerate}
    
  \item {The \emph{YJH} spectrum of} GJ\,758\,B is best fit with a spectral type {$\text{T}7.0 \pm 1.0$,} an effective temperature {$T_\text{eff} = 741 \pm 25$\,K} and surface gravity {$\log g = 4.3 \pm 0.5$\,dex (cgs)}, but a slightly later spectral type and lower $T_\text{eff}$, compatible with results of \citet{Thalmann2009}, \citet{Currie2010}, and \citet{Janson2011b}, cannot be excluded.

  \item A \changed{calculated $J-H$ color of {$-0.7 \pm 0.3$~mag} supports the found spectral type}, in comparison to field brown dwarfs, but again does not constrain it to exclude later spectral types.

  \item Combined with our derived effective temperature, evolutionary models suggest the companion has a mass $M = 40$--50\,$M_\mathrm{Jup}$, for an assumed age of 5--9\,Gyr.

  \item Molecular absorption features in the spectrum of GJ\,758\,B confirm the presence of methane in its atmosphere.

  \item An RV trend of the primary, GJ\,758, due to gravitational interaction with the companion, is predicted to be observable within a decade of regular monitoring, which together with astrometry could allow a model-independent dynamical mass to be derived from orbit analysis, and possibly make GJ\,758\,B the coolest substellar object to be used as a standard point of reference for mass and spectrum determination of T~dwarfs.

\end{enumerate}

More detailed astrometric analysis is being performed, and will be applied to extensive modeling of GJ\,758\,B's orbital motion, in order to constrain its orbital parameters. This will be presented in J.~Aguilar et al.\ (in preparation).

\acknowledgments
\textbf{Acknowledgments}
We are grateful for the financial, scientific, and technical support that made this research possible: R.N.\ was funded by the Swedish Research Council's International Postdoctoral Grant No.~637-2013-474. A portion of this work was supported by NASA Origins of the Solar System grant No.~NMO7100830/102190, and NASA APRA grant No.~08-APRA08-0117. E.R.\ acknowledges support from the National Science Foundation under Grant No.~1211568 and NASA Astrophysics Data Analysis Program (ADAP) award 11-ADAP11-0169. J.A.'s work was facilitated in part by a National Physical Science Consortium Fellowship and by stipend support from the Laboratory for Physical Sciences in College Park, Maryland. Part of the research was carried out at the Jet Propulsion Laboratory, California Institute of Technology, under a contract with the National Aeronautics and Space Administration. We also thank the Palomar mountain crew, especially Bruce Baker, Mike Doyle, Carolyn Heffner, John Henning, Greg van Idsinga, Steve Kunsman, Dan McKenna, Jean Mueller, Kajsa Peffer, Paul Nied, Joel Pearman, Kevin Rykoski, Carolyn Heffner, Jamey Eriksen, and Pam Thompson.

Facilities: \facility{Palomar Observatory, Hale(P3k,P1640)}.

%\bibliography{GJ758}
\bibliographystyle{apj.bst}

\end{document}